\documentclass[prd,aps,twocolumn,a4paper,floatfix,showpacs,nofootinbib]{revtex4-1}

\usepackage{graphicx,psfrag}
\usepackage{mathrsfs}
\usepackage{amsmath,amsfonts,amssymb}
\usepackage{multirow,enumerate}
\usepackage{comment,hyperref}
\usepackage{color}

\newcommand{\be}{\begin{equation}}
\newcommand{\ee}{\end{equation}}
\newcommand{\bea}{\begin{eqnarray}}
\newcommand{\eea}{\end{eqnarray}}
\newcommand{\bel}{\begin{align}}
\newcommand{\eel}{\end{align}}

\newcommand{\mean}[1]{\langle#1\rangle}

\def\p{\partial}

\def\GMc2{{\rm G M_{\odot} c^{-2}}}

\def\Qw{Q_{\hat{\omega}}}
\def\hw{\hat{\omega}}
\def\SASB{S\leftrightarrow S}

\usepackage{color}

\definecolor{cyan}{rgb}{0,0.9,0.9}
\definecolor{orange}{rgb}{0.9,0.5,0}
\definecolor{magenta}{rgb}{1,0,1}
\definecolor{purple}{rgb}{0.8,0.4,0.8}
\definecolor{gray}{rgb}{0.5,0.5,0.5}

\begin{document}

\title{Gravitational waves and mass ejecta from binary neutron star mergers:\\
  Effect of the stars' rotation}

 \author{Tim \surname{Dietrich}${}^{1}$}
 \author{Sebastiano \surname{Bernuzzi}$^{2}$}
 \author{Maximiliano \surname{Ujevic}$^{3}$}
 \author{Wolfgang \surname{Tichy}$^{4}$}
 \affiliation{${}^1$Max Planck Institute for Gravitational Physics, Albert Einstein Institute, D-14476 Golm, Germany}
 \affiliation{${}^2$ DiFeST, University of Parma, and INFN Parma  I-43124 Parma, Italy}  
 \affiliation{${}^3$Centro de Ci\^encias Naturais e Humanas, Universidade Federal do ABC, 09210-170, Santo Andr\'e, S\~ao Paulo, Brazil}
 \affiliation{${}^4$ Department of Physics, Florida Atlantic University, Boca Raton, FL 33431 USA}
\date{\today}

\begin{abstract} 
We present new (3+1) dimensional numerical relativity simulations of the binary
neutron star (BNS) mergers that take into account the NS spins. 
We consider different spin configurations, aligned or antialigned to the orbital
angular momentum, for equal and unequal mass BNS and for two equations
of state. All the simulations employ quasiequilibrium circular initial
data in the constant rotational velocity approach, i.e.~they are consistent
with Einstein equations and in hydrodynamical equilibrium. We study the NS rotation
effect on the energetics, the gravitational waves (GWs) and on the possible
electromagnetic (EM) emission associated to dynamical mass ejecta. 
For dimensionless spin magnitudes of $\chi\sim0.1$
we find that spin-orbit interactions and also
spin-induced--quadrupole deformations
affect the late-inspiral--merger dynamics. The latter is,
however, dominated by finite-size effects.
Spin (tidal) effects contribute to GW phase differences up to $\sim5$
($20$) radians accumulated during the last eight orbits to merger.
Similarly, after merger the collapse time of the remnant 
and the GW spectrogram are affected by the NSs rotation.
Spin effects in dynamical ejecta are clearly observed in unequal mass systems
in which mass ejection originates from the tidal tail of the companion.
Consequently kilonovae and other EM counterparts are affected by spins.
We find that spin aligned to the orbital angular momentum leads 
to brighter EM counterparts than antialigned spin 
with luminosities up to a factor of two higher.
\end{abstract}

\pacs{
  04.25.D-,     
  04.30.Db,   
  95.30.Lz,   
  97.60.Jd      
}

\maketitle

\nocite{Dietrich:2016hky}
\section{Introduction}
\label{sec:intro}

With the detection of the first gravitational wave (GW) 
signals GW150914~\cite{Abbott:2016blz} and GW151226~\cite{Abbott:2016nmj} 
the era of GW astronomy has begun. 
Beside black hole binaries, binary neutron stars (BNS) are one of the expected 
sources for future detections with the advanced GW interferometers~\cite{Aasi:2013wya}.
The theoretical modeling of the GW signal is crucial to support future GW
astronomy observation of BNS. BNS mergers are also expected to be
bright in the electromagnetic (EM) spectrum. Possible EM counterparts
of the GW signal are short gamma-ray bursts \cite{Paczynski:1986px,Eichler:1989ve,Soderberg:2006bn}, 
kilonovae~\cite{Tanvir:2013pia,Yang:2015pha,Jin:2016pnm,Metzger:2016pju} (also referred to as macronovae) 
and radio flares~\cite{Nakar:2011cw}. Detailed models of EM counterparts
will help the development of multimessenger astronomy.

Modeling BNS mergers requires to cover the entire
parameter space of BNSs, including the stars' rotational (spin) effects.  
Although observations suggest that most neutron stars (NSs) in binary systems 
have comparable individual masses $\sim 1.35 M_\odot$ and 
relatively small spins
\cite{Kiziltan:2013oja,Lattimer:2012nd}, this
conclusion might be biased by the small
number of observed BNS. The BNS parameter space could be much richer, 
in particular population synthesis models predict a wider range of masses and 
mass ratios~\cite{Dominik:2012kk,Dietrich:2015pxa}. 
Recent observations of compact binary systems with 
mass ratios of $q \approx 1.3$ suggest that BNSs with a significant
mass asymmetry can  exist~\cite{Martinez:2015mya,Lazarus:2016hfu}.
As far as spins are concerned, pulsar data indicate that NS can have a
significant rotation even in binary systems. Some of these NS in
binaries 
approach the rotational frequency of isolated milli-second pulsars.

For example, the NS in the binary system PSR J1807$-$2500B has a
rotation frequency of $239$Hz \cite{Lorimer:2008se,Lattimer:2012nd}, 
and one of the double pulsar components   
(PSR J0737$-$3039A) has rotational a
frequency of $44$Hz \cite{Burgay:2003jj}.
There is also evidence that dynamical capture and exchange interactions involving NSs
are a frequent occurrence in globular clusters~\cite{Verbunt:2013kka}; 
during this process exotic objects, 
such as double millisecond pulsars might form~\cite{Benacquista:2011kv}.

The only possibility to study the dynamics and waveforms in the time
period shortly before and after the merger of BNS systems is to
perform 
numerical relativity (NR) simulations that include general relativistic
hydrodynamics (GRHD). Despite the large progress of the field during
the last 10 years, spin effects in BNS mergers have been investigated
in few works. A main reason was the lack of consistent and realistic
initial data for the simulations, a crucial prerequisite for NR evolutions. 
General-relativistic quasi-equilibrium configurations of rotating NSs
of circular binary system can be now computed within the constant
rotational velocity (CRV) approach~\cite{Tichy:2011gw,Tichy:2012rp}.
These data are neither corotational nor irrotational, and permit, for
the first time, the NR-GW modeling
of realistic BNS sources with spins. (See Sec.~I of
\cite{Bernuzzi:2013rza} for a discussion). Alternative NR evolutions
of spinning BNS were presented
  in~\cite{Kastaun:2013mv,Tsatsin:2013jca,Kastaun:2014fna,East:2015vix},
  but employed constraint violating initial data.
Spinning BNS were also considered with a smooth particle hydrodynamics code
under the assumption of conformal flatness,
e.g.~\cite{Bauswein:2015vxa}.

Evolutions of CRV initial data have been considered 
in~\cite{Bernuzzi:2013rza,Dietrich:2015pxa,Tacik:2015tja}.
We have presented the first evolutions covering the last 3 orbits and 
postmerger for a 
BNS systems described by polytropic equations of state
\cite{Bernuzzi:2013rza}. That work proposed an analysis of the
conservative dynamics in terms of gauge-invariant curves of the binding
energy vs. angular momentum and a very preliminary analysis of the
spin effects on the waveform. 
In~\cite{Dietrich:2015pxa} we have made significant improvements in
the way we construct CRV initial data, which allows us to investigate
BNS mergers in an extended parameter space, and presented preliminary 
evolutions of generic mergers (i.e.~with precession). 
Ref.~\cite{Tacik:2015tja} presented an independent implementation of
CRV initial data and preliminary evolutions, but did not cover the
final merger and postmerger phases. 

Several important questions remain open. 
A detailed understanding of the role of spin interactions will
be fundamental for building analytical models of the inspiral--merger
phase. Thus, it is important to further explore the BNS dynamics with 
long simulations and spanning a larger parameter space than previously
considered. The influence of the NS spins on the GW phase evolution
during the last orbits and up to merger is not fully understood but is
very relevant for GW data analysis \cite{Brown:2012qf,Agathos:2015uaa}. Understanding
the spin influence on the merger remnant might be relevant for both GW and
EM observations. Also, the role
of the NSs rotation on the dynamical ejecta 
and on the EM counterparts has not been explored.

In this article, we investigate rotational (spin) effects in
multi-orbit BNS merger simulations with
different mass-ratio and propose the first answers to the questions above. 
Our simulations cover $\sim12$ orbits to merger and
postmerger for mass ratios $q=1,1.25,1.5$, two different equations of
state (EOSs), and spin aligned or anti-aligned to the orbital angular momentum.
These simulations are the first of their kind, and will support the
development of analytical models of the GWs and of the EM emission
from merger events.
This paper extends the results of Ref.~\cite{Dietrich:2016hky}
(hereafter Paper I) that was limited to irrotational
configurations and focused on the effect of the mass ratio. 
Our goal is to cover a significant part of the BNS
parameter space. 

The article is structured as follows: 
In Sec.~\ref{sec:methods}, we describe briefly
the numerical methods and some analysis tools. 
In Sec.~\ref{sec:config} we present the configurations 
employed in this work. 
Section~\ref{sec:dynamics} summarizes the dynamics 
of the merger process, where 
the spin evolution of the individual stars and the energetics during the 
inspiral and post-merger are discussed. 
In Sec.~\ref{sec:ejecta}-\ref{sec:EM} dynamical ejecta, 
the GW signal, and possible electromagnetic (EM) counterparts 
are studied. We conclude in Sec.~\ref{sec:conclusion}. 

Throughout this work we use geometric units, setting $c=G=M_\odot=1$,
though we will sometimes include $M_\odot$ explicitly or quote values
in cgs units for better understanding and astrophysical interpretation.  
Spatial indices are denoted by Latin letters running from 1 to 3 and 
Greek letters are used for spacetime indices running from 0 to 3.

\section{Simulation methods}
\label{sec:methods}

\subsection{Initial configurations}

Our initial configurations are constructed with the pseudospectral  
SGRID code~\cite{Tichy:2006qn,Tichy:2009yr,Tichy:2009zr,Dietrich:2015pxa}.
We use the conformal thin sandwich equations~\cite{Wilson:1995uh,Wilson:1996ty,York:1998hy}
together with the CRV approach~\cite{Tichy:2011gw,Tichy:2012rp} to solve 
the constraint equations. 
We construct quasi-equilibrium configuration in quasi-circular orbits,
assuming a helical Killing vector. We follow exactly the same setup as
in Paper~I to which we refer for more details. 

In order to construct BNS with different spins the approach
of~\cite{Tichy:2009zr,Dietrich:2015pxa} is adopted.
The CRV method does not allow to prescribe the spin (or the
dimensionless spin) directly; only the rotational part of the
four-velocity can be specified as free data. 
We use Eq.~(C3) of Ref.~\cite{Dietrich:2015pxa} to obtain an estimate
for the four-velocity corresponding to a given
dimensionless spin of $\chi=0.1$. Once the rotational velocity is
fixed, we compute a single NS with the same baryonic mass as the one
in the binary and measure its ADM angular momentum. This gives the
dimensionless spin of one component of the binary. The procedure is
repeated for the other component.

For binary configurations in quasi-equilibrium, the described procedure
gives consistent results between the ADM angular momentum of the spinning 
and irrotational BNS. In particular the difference between the 
$J_{\rm ADM}$ of the spinning and irrotational BNS is 
consistent with the sum of the spin estimates,  $\Delta J_{\rm
  ADM}\sim(S^A+S^B)$, up to $10^{-2}$; fractional errors are always
$\lesssim 0.3 \%$.  
Those small differences might also be caused by small differences 
in the initial orbital frequency.

The properties of the initial BNS configuration are summarized in
Tab.~\ref{tab:config}, and discussed in more detail in
Sec. \ref{sec:config}. 

\begin{tiny}
\begin{table*}[t]
  \centering    
  \caption{BNS configurations. 
    The first column defines the configuration name. 
    Next 9 columns describe the physical properties of the single stars: 
    the EOS, the gravitational masses of the individual stars $M^{A,B}$, 
    the baryonic masses of the individual stars $M_{b}^{A,B}$, the stars' spins
    $S^{A,B}$, and dimensionless spins $\chi^{A,B}$. 
    The last 5 columns define the tidal coupling constant $\kappa_2^T$,
    the mass-weighted spin $\chi_{mw}$, 
    the initial GW frequency $M \omega_{22}^0$, 
    the ADM-Mass $M_{\rm ADM}$, 
    and the ADM-angular momentum $J_{\rm ADM}$.}
    \setlength{\tabcolsep}{1pt}
  \begin{tabular}{c|l|ccccccccc|ccccc}        
   & Name & EOS & $M^A$ & $M^B$ & $M_b^A$ & $M_b^B$ & $S^A$ & $S^B$ & $\chi^A$ & $\chi^B$ & $\kappa_2^T$ & $\chi_{\rm mw}$ & $M \omega_{22}^0$ & $ M_{ADM}$ &  $J_{ADM}$ \\
     \hline
     \hline
      \multirow{4}{*}{\rotatebox[origin=c]{90}{\textbf{$q=1.00$}}}
   &  ALF2-137137$^{(00)}$                  & ALF2 & 1.375008 & 1.375008 & 1.518152 & 1.518152 & 0.0000 & 0.0000  & 0.000 & 0.000  & 125 & 0.000 & 0.0360 & 2.728344 & 8.1200 \\  
   &  ALF2-137137$^{(\uparrow \uparrow)}$   & ALF2 & 1.375516 & 1.375516 & 1.518152 & 1.518152 & 0.1936 & 0.1936  & 0.102 & 0.102  & 125 & 0.102 & 0.0360 & 2.729319 & 8.4811 \\   
   &  ALF2-137137$^{(\uparrow \downarrow)}$ & ALF2 & 1.375516 & 1.375516 & 1.518152 & 1.518152 & 0.1936 & -0.1936 & 0.102 & -0.102 & 125 & 0.000 & 0.0360 & 2.729333 & 8.1240 \\
   &  ALF2-137137$^{(\uparrow 0)}$          & ALF2 & 1.375516 & 1.375008 & 1.518152 & 1.518152 & 0.0000 & 0.1936  & 0.102 & 0.000  & 125 & 0.051 & 0.0360 & 2.728816 & 8.2997 \\
   \hline       \multirow{4}{*}{\rotatebox[origin=c]{90}{\textbf{$q=1.00$}}}
   &  H4-137137$^{(00)}$                    & H4   & 1.375006 & 1.375006 & 1.498528 & 1.498528 & 0.0000 & 0.0000  & 0.000 & 0.000  & 188 & 0.000 & 0.0348 & 2.728211 & 8.0934 \\
   &  H4-137137$^{(\uparrow \uparrow)}$     & H4   & 1.375440 & 1.375440 & 1.498528 & 1.498528 & 0.1892 & 0.1892  & 0.100 & 0.100  & 188 & 0.100 & 0.0348 & 2.729056 & 8.4508 \\
   &  H4-137137$^{(\uparrow \downarrow)}$   & H4   & 1.375440 & 1.375440 & 1.498528 & 1.498528 & 0.1892 & -0.1892 & 0.100 & -0.100 & 188 & 0.000 & 0.0349 & 2.729067 & 8.0983 \\
   &  H4-137137$^{(\uparrow 0)}$            & H4   & 1.375440 & 1.375006 & 1.498528 & 1.498528 & 0.1892 & 0.000   & 0.100 & 0.000  & 188 & 0.050 & 0.0348 & 2.728643 & 8.2711 \\      
     \hline
     \hline     \multirow{4}{*}{\rotatebox[origin=c]{90}{\textbf{$q=1.25$}}}
   &  ALF2-122153$^{(00)}$                  & ALF2 & 1.527790 & 1.222231 & 1.707041 & 1.334040 & 0.0000 & 0.0000  & 0.000 & 0.000  & 127 & 0.000 & 0.0357 & 2.728212 & 7.9556 \\
   &  ALF2-122153$^{(\uparrow \uparrow)}$   & ALF2 & 1.528484 & 1.222602 & 1.707041 & 1.334040 & 0.2430 & 0.1521  & 0.104 & 0.102  & 127 & 0.103 & 0.0357 & 2.729255 & 8.3300 \\ 
   &  ALF2-122153$^{(\uparrow \downarrow)}$ & ALF2 & 1.528484 & 1.222602 & 1.707041 & 1.334040 & 0.2430 & -0.1521 & 0.104 & -0.102 & 127 & 0.013 & 0.0358 & 2.729256 & 8.0479 \\
   &  ALF2-122153$^{(\uparrow 0)}$          & ALF2 & 1.528484 & 1.222231 & 1.707041 & 1.334040 & 0.2430 & 0.0000  & 0.104 & 0.000  & 127 & 0.058 & 0.0357 & 2.728907 & 8.1895 \\  
   \hline       \multirow{4}{*}{\rotatebox[origin=c]{90}{\textbf{$q=1.25$}}}
   &  H4-122153$^{(00)}$                    & H4   & 1.527789 & 1.222228 & 1.683352 & 1.318080 & 0.0000 & 0.0000  & 0.000 & 0.000  & 193 & 0.000 & 0.0349 & 2.728675 & 8.0248 \\ 
   &  H4-122153$^{(\uparrow \uparrow)}$     & H4   & 1.528365 & 1.222546 & 1.683352 & 1.318080 & 0.2329 &  0.1499 & 0.100 & 0.100  & 193 & 0.100 & 0.0349 & 2.729567 & 8.3899 \\      
   &  H4-122153$^{(\uparrow \downarrow)}$   & H4   & 1.528365 & 1.222546 & 1.683352 & 1.318080 & 0.2329 & -0.1499 & 0.100 & -0.100 & 193 & 0.011 & 0.0349 & 2.729585 & 8.1135 \\  
   &  H4-122153$^{(\uparrow 0)}$            & H4   & 1.528365 & 1.222228 & 1.683352 & 1.318080 & 0.2329 &  0.0000 & 0.100 & 0.00   & 193 & 0.056 & 0.0349 & 2.729250 & 8.2491 \\
   \hline
   \hline       \multirow{4}{*}{\rotatebox[origin=c]{90}{\textbf{$q=1.50$}}}
   &  ALF2-110165$^{(00)}$                   & ALF2 & 1.650015 & 1.100016 & 1.862057 & 1.189870 & 0.0000 & 0.0000  & 0.000 & 0.000  & 133 & 0.000 & 0.0356 & 2.728542 & 7.6852 \\  
   &  ALF2-110165$^{(\uparrow \uparrow)}$   & ALF2 & 1.650924 & 1.100296 & 1.862057 & 1.189870 & 0.2919 & 0.1223  & 0.107 & 0.101  & 133 & 0.105 & 0.0355 & 2.729669 & 8.0732 \\     
   &  ALF2-110165$^{(\uparrow \downarrow)}$ & ALF2 & 1.650924 & 1.100296 & 1.862057 & 1.189870 & 0.2919 & -0.1223 & 0.107 & -0.101 & 133 & 0.024 & 0.0355 & 2.729677 & 7.8475 \\
   &  ALF2-110165$^{(\uparrow 0)}$          & ALF2 & 1.650924 & 1.100016 & 1.862057 & 1.189870 & 0.2919 & 0.0000  & 0.107 & 0.000  & 133 & 0.064 & 0.0355 & 2.729404 & 7.9599 \\
   \hline       \multirow{4}{*}{\rotatebox[origin=c]{90}{\textbf{$q=1.50$}}}
   &  H4-110165$^{(00)}$                    & H4   & 1.650017 & 1.100006 & 1.834799 & 1.176579 &  0.0000 & 0.0000 & 0.000 & 0.000  & 209 & 0.000 & 0.0350 & 2.729385 & 7.81991 \\
   &  H4-110165$^{(\uparrow \uparrow)}$     & H4   & 1.650752 & 1.100242 & 1.834799 & 1.176579 &  0.2745 & 0.1204 & 0.101 & 0.099  & 209 & 0.100 & 0.0350 & 2.730283 & 8.18713 \\
   &  H4-110165$^{(\uparrow \downarrow)}$   & H4   & 1.650752 & 1.100242 & 1.834799 & 1.176579 &  0.2745 & 0.1204 & 0.101 & -0.099 & 209 & 0.021 & 0.0350 & 2.730267 & 7.96085 \\
   &  H4-110165$^{(\uparrow 0)}$            & H4   & 1.650752 & 1.100006 & 1.834799 & 1.176579 &  0.2745 & 0.0000 & 0.101 & 0.000  & 209 & 0.061 & 0.0350 & 2.730050 & 8.07357 \\
\hline \hline
  \end{tabular}
 \label{tab:config}
\end{table*}
\end{tiny}

\subsection{Evolutions}

Dynamical simulations are performed with the BAM 
code~\cite{Thierfelder:2011yi,Brugmann:2008zz,Dietrich:2015iva}, 
employing the Z4c scheme~\cite{Bernuzzi:2009ex,Hilditch:2012fp} and the 1+log and 
gamma-driver conditions for the 
gauge system~\cite{Bona:1994a,Alcubierre:2002kk,vanMeter:2006vi}.
The GRHD equations are solved in
conservative form by defining Eulerian conservative variables from the
rest-mass density $\rho$, pressure $p$, internal energy $\epsilon$, and
3-velocity $v^i$ with a high-resolution-shock-capturing 
method~\cite{Thierfelder:2011yi} based on primitive reconstruction
and the Local-Lax-Friedrichs central scheme for the numerical
fluxes. The GRHD system is closed by an EOS.
We work with two EOSs modeled as piecewise polytropic 
fits~\cite{Read:2008iy,Dietrich:2016hky} 
and include thermal effects with an additive pressure contribution 
$p_{\rm th} = (\Gamma_{\rm th}-1)\rho\epsilon$~\cite{Shibata:2005ss,Bauswein:2010dn}
setting $\Gamma_{\rm th}=1.75$.
The Berger-Oliger algorithm is employed for the time stepping~\cite{Berger:1984zza} and we 
make use of an additional refluxing algorithm to enforce mass conservation 
across mesh refinement boundaries~\cite{Berger:1989,East:2011aa}
as in previous works~\cite{Dietrich:2014wja,Dietrich:2015iva,Dietrich:2016hky}. 
Restriction and prolongation between the refinement levels is performed with 
an average scheme and 2nd order essentially non-oscillatory
scheme, respectively.

We employ the same grid setup as the ``shell'' setup in~Paper~I, 
i.e.~the numerical domain is made of a hierarchy of cell-centered nested
Cartesian grids, where the outermost level is substituted by a
multipatch (cubed-sphere) grid~\cite{Ronchi:1996,Thornburg:2004dv,Pollney:2009yz,Hilditch:2012fp}.
In total we have used 4 different grid setups summarized in Tab.~\ref{tab:grid}.

\begin{table}[t]
  \centering    
  \caption{Grid configuration:
    name, EOS, finest grid spacing $h_{L-1}$, radial resolution inside the
    shells $h_r$, number of points $n$ ($n^{mv}$) in the fixed (moving) levels, 
    radial point number $n_r$ and azimuthal number of points $n_\theta$ in the shells, 
    inradius up to which GRHD equations are solved $r_1$, and the outer boundary $r_b$.}
  \begin{tabular}{l|l|cccccccc}        
   Name & EOS & $h_{L-1}$ & $h_{r}$ & $n$ & $n^{mv}$ & $n_r$ & $n_\theta$ & $r_1$ & $r_b$ \\
     \hline
     \hline
  R1    & ALF2 & 0.250  & 8.00 & 128 & 64  & 128 & 64 & 572 & 1564 \\  
  R2    & ALF2 & 0.167  & 5.33 & 192 & 96  & 192 & 96 & 552 & 1555 \\  
\hline
  R1    & H4   & 0.250  & 8.00 & 128 & 72  & 128 & 64 & 572 & 1564 \\  
  R2    & H4   & 0.167  & 5.33 & 192 & 108 & 192 & 96 & 552 & 1555 \\ 
\hline \hline
  \end{tabular}
 \label{tab:grid}
\end{table}

\subsection{Simulation analysis} 
 
Most of our analysis tools were summarized 
in Sec.~III of Paper~I. They include the computation of the ejecta
quantities, the disk masses,  
the entropy indicator, the amount of mass transfer during the inspiral, 
and the way we extract GWs. 
Here, we extend the analysis tools by including a quasi-local measure of the 
spin of the NSs. Following a similar approach as in
Refs.~\cite{Tacik:2015tja}, 
we evaluate the surface integral
\begin{equation}
\label{eq:quasi_local_spin}
 S^i \approx \frac{1}{8\pi} 
 \int_{r_s} \text{d}^2 x \sqrt{\gamma} \left( \gamma^{k j} K_{l k} - \delta_{l}^{j} K \right) 
 n_{j} \varphi^{li}, 
\end{equation}
on coordinate spheres with radius $r_S$ around the NSs. 
$\varphi^{li} = \epsilon^{l i k } x_k $ defines the approximate rotational Killing vectors 
in Cartesian coordinates ($\varphi^{l1},\varphi^{l2},\varphi^{l3}$), 
$K_{ij}$ denotes the extrinsic
curvature, $\gamma^{ij}$ the inverse 3-metric,  
and $n_i = (x_i-x_i^{\rm NS})/r$ the normal vector with respect to the
center of the NS. 
The center is given by the minimum of the lapse inside
the NS\footnote{While 
this paper has been written also~\cite{Kastaun:2016private} implemented 
the exactly same method as proposed here to measure the spin of the 
single NSs during BNS inspirals. Both implementations 
have been compared and give similar results.}. 
Differently from \cite{Tacik:2015tja} we do not determine the center of the 
coordinate sphere by the maximum density and we do not use comoving coordinates in our
simulations.

Let us discuss the interpretation of 
Equation~\eqref{eq:quasi_local_spin}.
For equilibrium rotating NS
spacetimes, Eq.~\eqref{eq:quasi_local_spin} in the limit
$r_S\to\infty$ reproduces the ADM angular momentum of the (isolated) NS,
see Appendix~\ref{app:quasi_local_single}. 
In dynamical BNS evolutions, Eq.~\eqref{eq:quasi_local_spin} allows us
to measure the spin evolution and spin direction. We stress that, in
the BNS case, the spin measure has some caveats:
(i) no unambiguous spin definition of a single object inside a binary
    system exists in general relativity;
(ii) Eq.~\eqref{eq:quasi_local_spin} is evaluated in the strong-field region although 
     it is only well defined at spatial infinity; 
(iii) the $r_S$ spheres are gauge dependent.

\section{BNS Configurations}
\label{sec:config}

We consider BNS configurations with fixed total mass of
$M=M^A+M^B=2.75M_\odot$, and vary EOS, mass-ratio, and the spins. The
spins are always aligned or antialigned to the orbital angular
momentum.  
The EOSs are ALF2 and H4; both support masses of isolated NSs above
$2M_\odot$ and are compatible with current astrophysical constraints. 
We vary the mass ratio, 
\begin{equation}
q := \frac{M^A}{M^B} \geq1 \ , 
\end{equation}
spanning the values $q=(1.0,1.25,1.5)$. 
For every EOS and $q$, we consider four different spin configurations:
\begin{enumerate}
 \item[$(00)$] none of the stars is spinning; 
 \item[$(\uparrow \uparrow)$] both spins are aligned with the orbital momentum; 
 \item[ $(\uparrow \downarrow)$] the spin of star A is aligned, the other
     star spin is anti-aligned; 

 \item[$(\uparrow 0)$] the spin of star A is aligned to the orbital angular
     momentum and the other NS is 
     considered to be irrotational, 
\end{enumerate}
where the dimensionless spin magnitude 
\be
\chi := \frac{S}{M^2}
\ee
of each star is either $\chi=0$ or $\chi\sim0.1$. 
The properties of the considered BNSs are summarized in Tab.~\ref{tab:config}. 

A BNS configuration is determined by its EOS, individual masses (or mass
ratio), and the two spins. Focusing on the GWs, we parametrize this
configuration space as follows. 
%
Spin effects are described by the mass-weighted spin combination, 
\begin{equation}
 \chi_{\rm mw} := \frac{M^A \chi^A + M^B \chi^B}{(M^A+M^B)} \ , \label{eq:chimw}
\end{equation}
which is used for phenomenological waveforms models and during GW searches, 
e.g.~\cite{Ajith:2009bn,TheLIGOScientific:2016pea}.
The mass-weighted spin is related to the effective spin $\chi_{\phi}$, 
which captures the leading order spin effects of the phase evolution via
\begin{equation}
\chi_{\phi} = \chi_{\rm mw} - \frac{38\nu}{113} (\chi^A + \chi^B)\ , 
\end{equation}
with the symmetric mass ratio $\nu = M^A M^B/ (M^A+M^B)^2$.
For the setups presented here $\chi_{\rm mw} \approx \chi_{\rm eff}$, 
which is the reason why we restrict us to the more commonly used $\chi_{\rm mw}$. 

Most of the NS structure and EOS information is encoded in the 
tidal polarizability coefficient~\cite{Bernuzzi:2014kca,Bernuzzi:2015rla}
\begin{equation}
\kappa^T_2 := 2 \left( \frac{q^4}{(1+q)^5} \frac{k_2^A}{C_A^5}  +
\frac{q}{(1+q)^5} \frac{k_2^B}{C_B^5} 
\right)  \ 
\label{eq:kappa}
\end{equation}
that describes at leading order the NSs' tidal
interactions. $\kappa^T_2$ depends on 
the EOS via the quadrupolar dimensionless Love number $k_2$ of isolated spherical
star configurations, e.g.~\cite{Damour:2009vw}, 
and the compactness $C$ of the irrotational stars (defined as the ratio of the
gravitational mass in isolation with the star's proper radius). 
%
As a further parameter we choose the mass ratio, since  
the dynamics of nonspinning black hole binary is entirely
described by $q$.

The 3D parametrization $(q,\chi_{\rm mw},\kappa_2^T)$ is a (possible) minimal
choice for the description of BNS GWs. The binary total
mass $M$, in particular, scales trivially in absence of tides and 
its dependency in the tidal waveform is hidden in the $\kappa^T_2$, to leading order.
It should be noted, however, that $(q,\chi_{\rm mw},\kappa_2^T)$ are
not independent variables and some degeneracies exists [$\kappa_2^T$ and $\chi_{\rm mw}$
depend on $q$ for instance]. Furthermore, we note that the intrinsic
NS rotation can also influence tidal effects during the evolution. 
In this work, we use for consistency $(q,\chi_{\rm
  mw},\kappa_2^T)$ to study the parametric dependency of other
quantities than GWs, like ejecta and EM luminosity.

\begin{figure}[t]
  \includegraphics[width=0.5\textwidth]{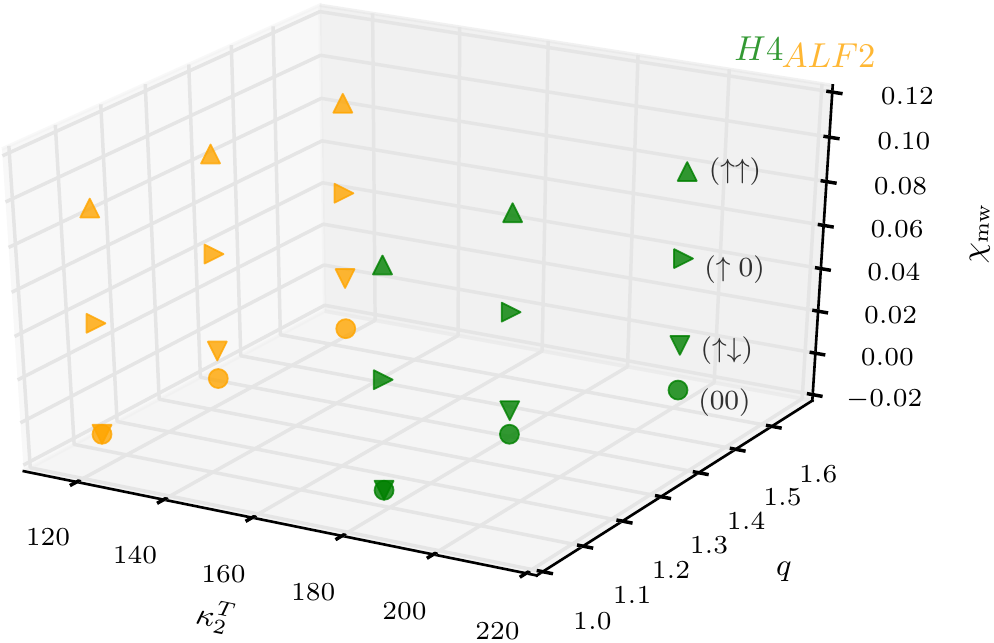}
  \caption{The $(q,\chi_{\rm mw},\kappa_2^T)$ parameter space
    coverage. 
    Different colors refer to the EOS ALF2 (orange) and H4 (green). 
    Different markers correspond to different spin configurations: 
    circles $(00)$, triangles pointing down ($\uparrow \downarrow$),
    triangles pointing right $(\uparrow 0)$, and 
    triangles pointing up $(\uparrow \uparrow)$.}
  \label{fig:param}
\end{figure}

The $(q,\chi_{\rm mw},\kappa_2^T)$ parameter space coverage of our
work is shown in Fig.~\ref{fig:param}. 
In total we are considering 24 BNSs. The irrotational configurations
were already presented in~Paper~I, but 36 new
simulations were performed for the scope of this paper.
Every configuration is simulated with two
different resolutions R1 and R2, see Tab.~\ref{tab:grid}. 
This allows us to place error bars on our results, which will be
conservatively estimated as the difference between the resolutions,
see~\cite{Bernuzzi:2011aq,Bernuzzi:2016pie} for a detailed analysis.  

\section{Dynamics}
\label{sec:dynamics}

\subsection{Qualitative discussion}
\label{sec:Qualitative_analysis} 

  \begin{figure*}[t]
  \includegraphics[width=1\textwidth]{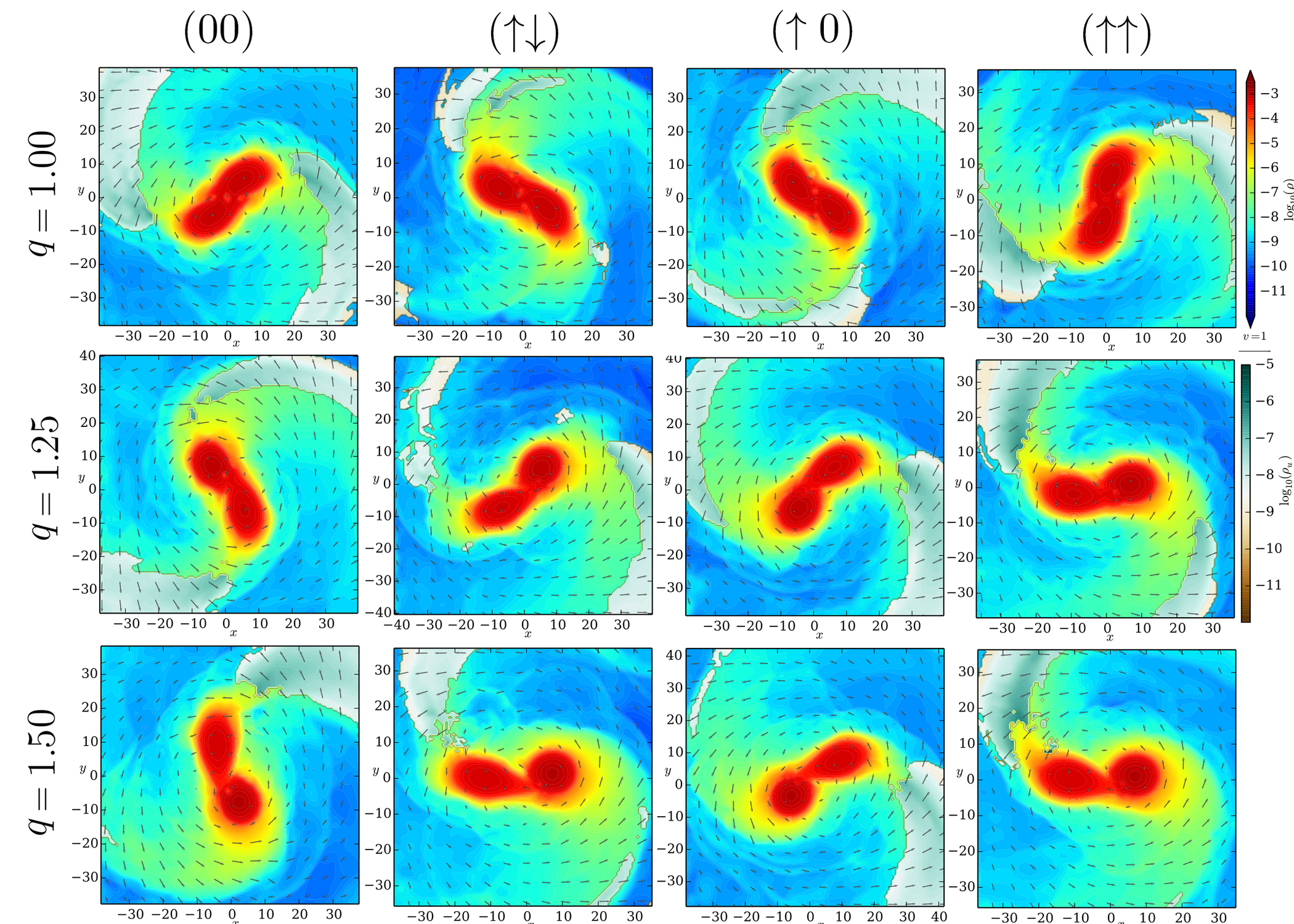}
  \caption{Rest-mass density profile inside the orbital plane for simulations employing the H4 EOS and using 
           the R2 grid setup. The snapshots represent the moments of merger. 
           The panels refer to (from top to bottom) mass ratios $q=1.00,q=1.25,q=1.50$ and 
           (from left to right) spin configurations $(00),(\uparrow \downarrow),(\uparrow 0),(\uparrow \uparrow)$.
           The rest-mass density $\rho$ is shown on a logarithmic scale from blue to red. 
           The rest-mass density of unbound material is colored from brown to green.
           Most material gets ejected from the tidal tails of the NS
           inside the orbital plane. } 
  \label{fig:2d_rho_H4}
  \end{figure*} 

Our simulations span $N_\text{orb}\sim10-12$ orbits (20-24 GW cycles)
to merger, the number of orbits increases (decreases) for spin
aligned (antialigned) to the 
orbital angular momentum. In this regime, spin effects typically
contribute up to $\Delta N_{Spin}\sim\pm1$ orbits. The spin effect 
is comparable to the effect of the EOS variation and of the mass
ratio, $\Delta N_{\rm EOS}\sim\Delta N_{q}\sim \Delta N_{Spin}$. BNS with
stiffer EOS and/or larger $q$ take fewer orbits to merge for a 
fixed initial GW
frequency\footnote{Recall that for our configurations the H4 EOS is
  stiffer than the ALF2 EOS.}. 

Figure~\ref{fig:2d_rho_H4} shows the rest-mass density profile inside the
orbital plane for the configurations employing  
the H4 EOS with resolution R2. The snapshots are taken at the moment of merger, 
i.e.,~at the time where the amplitude of the GW has its maximum. 

Although the initial orbital frequency is almost identical for all systems with 
the same mass-ratio and EOS,
cf.~Tab.~\ref{tab:config}, the ``orbital phase'' at the moment of
merger differs due to the spin of the individual stars. 
In general, if a NS has spin aligned to the orbital angular momentum,
the binary dynamics is less bound leading to a slower phase evolution  
with respect to the irrotational case. Contrary, if a NS has anti-aligned spin
the binary is more bound leading to a faster phase evolution
and an earlier moment of merger,  
i.e.~at lower frequencies (see Sec.~\ref{sec:GW}). 
This {\it spin-orbit} (SO) effect has a solid analytical
basis~\cite{Damour:2001tu}, 
and was already reported in both BBH
simulations~\cite{Campanelli:2006uy} (``orbital hang-up'' effect) and in BNS
setups~\cite{Kastaun:2013mv,Tsatsin:2013jca,Bernuzzi:2013rza,Dietrich:2015pxa}. 

In the BNS configurations with $(\uparrow \downarrow)$ and
equal masses ($q=1$) the SO effect is zero at leading order, 
cf.~Eq.~\eqref{Hso} and discussion below. 
Notably in these cases, the effects of the {\it spin-spin} interactions
(SS) are observed in our simulations. Comparing the irrotational BNS
$(00)$ with the $(\uparrow \downarrow)$ configuration, 
the latter has a faster phase evolution, i.e.~merges at lower frequencies.

After merger, the simulations are continued for about $\sim 30$ms. 
All the BNS considered in this work form a hypermassive neutron star
(HMNS). The presence of spin
influences the angular momentum of the remnant
HMNS. Configurations with $(\uparrow \uparrow)$, for example,
have additional angular momentum support and the 
HMNS has a longer lifetime. Spin effects influence the HMNS's
rotation law and its dynamical evolution [see 
Sec.~\ref{sec:GW:postmerger} for a detailed discussion]. 
Overall, spin effects are observed in the remnant and ejecta, but
better resolved during the early part of the simulations.

\subsection{Spin Evolution}
\label{sec:Spin-Evolution} 

\begin{figure}[t]
  \includegraphics[width=0.5\textwidth]{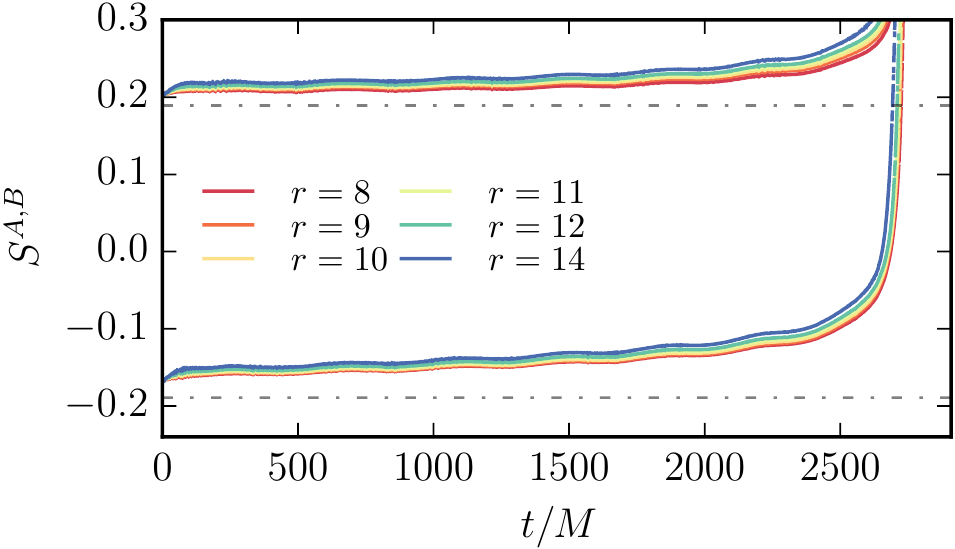}
  \caption{Quasi-local measurement of the individual spin of the two NSs
           for H4-137137$^{(\uparrow \downarrow)}$. 
           The gray dashed dotted lines in the diagram represent the value 
           computed from the initial data solver and given in Tab.~\ref{tab:config}. 
           Different colors refer to different radii of the coordinate spheres. 
           The increasing quasi-local spin is most likely caused by the choice of  
           the center of the integration surface and by using a coordinate sphere not taking tidal deformations 
           into account.}
  \label{fig:quasi_local}
  \end{figure}

The evolution of the quasi-local spin computed by
Eq.~\eqref{eq:quasi_local_spin}, is shown in
Fig.~\ref{fig:quasi_local} for the representative case
H4-137137$^{(\uparrow \downarrow)}$. 
We find that, within our uncertainties, the spins magnitudes remain
roughly constant up to the actual collision of the two stars. 
When the two stars finally merge, there is a single surface integral,
and Eq.~\eqref{eq:quasi_local_spin} estimates the orbital angular
momentum of the merger remnant.
Our results are consistent with what was observed in~\cite{Tacik:2015tja},
although the latter do not extend to merger.
They are also consistent with BBH simulations in which spins remain roughly constant
up to the formation of a common horizon
\cite{Lovelace:2010ne,Scheel:2014ina,Ossokine:2015vda}. 
However, comparing with \cite{Tacik:2015tja} our results have larger
uncertainties, whose origin we discuss in the following.

Fig.~\ref{fig:quasi_local} shows that, during the evolution, the spin
magnitude of the NS with spin aligned to the orbital angular momentum
seems to be larger by $\sim 20\%$ compared to the other. This happens
despite the fact that the  rotational velocities (and initial spin
values) are initially of the same magnitude. A similar effect was
shown in~\cite{Tacik:2015tja}, but it is more pronounced in our setup. \\

We argue this is caused by the fact that we are 
using coordinate spheres in a non comoving 
coordinate system. As a result, our setup does not capture accurately the 
(approximate) rotational symmetry around each star, the latter being
numerically entangled with the orbital motion.
The spin magnitudes are, consequently, overestimated.
For the same reason we observe a drift of the spin magnitude that
increases the closer the merger is. We believe this effect is partially
related to numerical accuracies.
Note finally that for irrotational BNSs we measure a residual spin
$S\sim 10^{-2}$. The value is consistent with the accuracy level of
the initial data, see also our results on isolated stars in
App.~\ref{app:quasi_local_single}. 

\subsection{Energetics}
\label{sec:dyn:inspiral}

  \begin{figure*}[t]
  \includegraphics[width=1\textwidth]{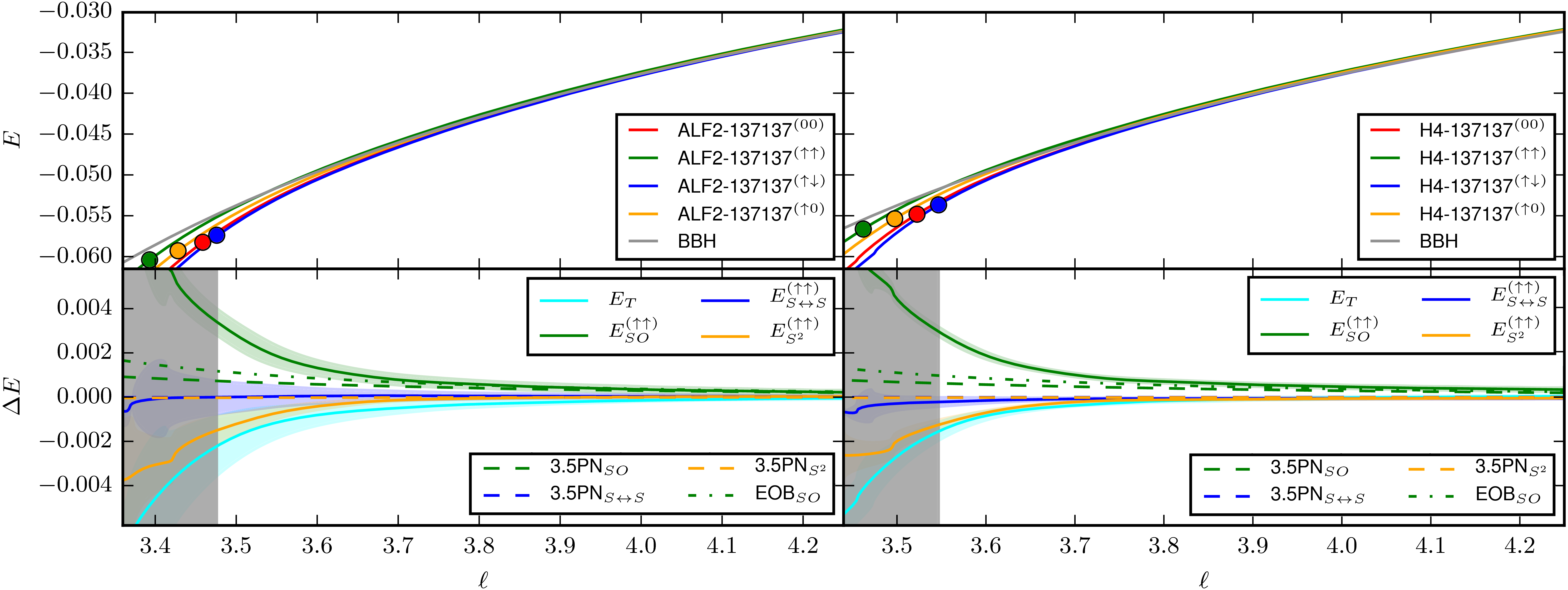}
  \caption{Binding energy vs.~specific angular momentum curves 
           $E(\ell)$ for the equal mass configurations (upper panels). 
           The circles mark the moment of merger for all configurations. 
           We also include a non-spinning BBH configuration from the public SXS 
           catalog, see text for more details. 
           The bottom panels show the individual contributions to the binding energy, 
           Eq.~\eqref{eq:Eb_ansatz}, we present the $SO$ (green), $S^2$ (orange), $\SASB$ (blue)
           configurations obtained from the NR data as solid lines. For those 
           contributions we also include 3.5PN (App.~\ref{app:PN_Ej}) estimates as dashed lines. 
           The tidal contributions are shown as cyan lines. 
           We also include the SO contribution from the EOB model of~\cite{Nagar:2015xqa}.
           We mark the difference between resolution R2 and R1 as a colored shaded region. 
           The vertical gray areas correspond to the merger of the ALF2-137137$^{(\uparrow \downarrow)}$ (left)
           and the H4-137137$^{(\uparrow \downarrow)}$ configuration (right).}
  \label{fig:Ej_H137137}
  \end{figure*} 

  \begin{figure}[t]
  \includegraphics[width=0.5\textwidth]{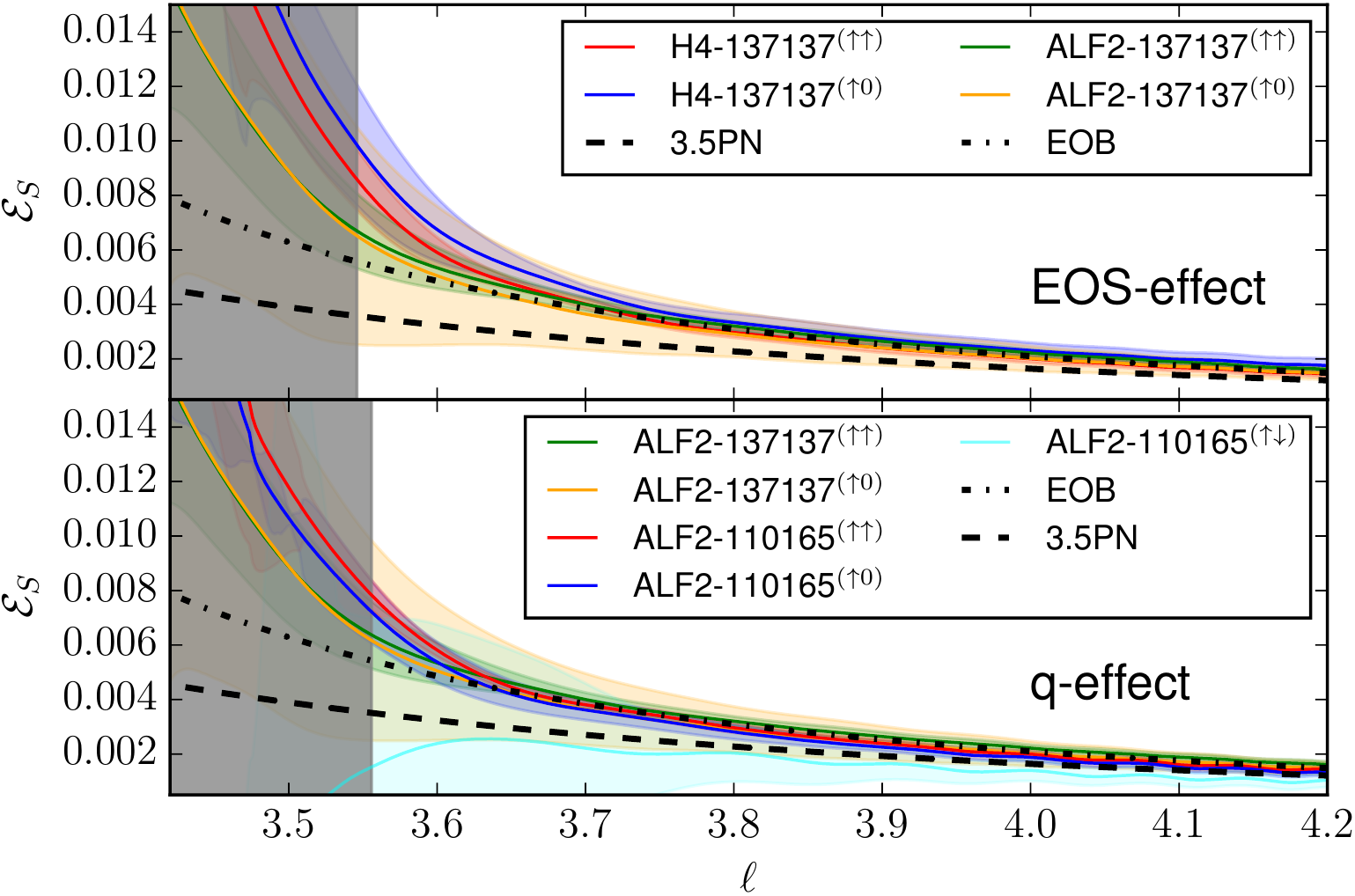}
  \caption{Quantity $\mathcal{E}_{Spin}$ for different EOS and the same mass ratio ($q=1.0$) in the top panel and 
           for the same EOS (ALF2), but mass ratios $q=1.0$ and $q=1.5$ in the bottom panel. 
           The colored shaded regions mark the difference between the results for resolution R2 and R1. 
           Notice that for an unequal mass merger also the $(\uparrow \downarrow)$ configuration 
           contains SO-contributions and therefore could be included 
           for $q =1.5$ although the late time behavior 
           is dominated by the SS-interactions.
           The vertical gray areas correspond to the moment of merger 
           for H4-137137$^{(\uparrow 0)}$ (upper panel) and ALF2-110165$^{(\uparrow \downarrow)}$ 
           (lower panel).
           We also include the 3.5PN SO contribution as a dashed line 
           and EOB $\mathcal{E}_{Spin}$ as present in the 
           EOB model of~\cite{Nagar:2015xqa} as a dashed dotted line.}
  \label{fig:Ebell_H4}
  \end{figure}  

We now discuss the BNS dynamics at a quantitative level by considering the
gauge invariant curves of the binding energy vs.~orbital angular 
momentum~\cite{Damour:2011fu} as well as the binding energy and
angular momentum dependency on the orbital frequency.

The specific binding energy is given by
\begin{equation}
\label{eq:Eb}
E_b = \frac{1}{\nu} \left[ \frac{M_{\rm ADM}(t=0)- \mathcal{E}_{\rm
      rad}}{M}-1 \right] \ , 
\end{equation}
where $\mathcal{E}_{\rm rad}$ is the energy emitted via GWs, as 
computed from the simulations (cf. Sec.~V of Paper I).
The specific and dimensionless orbital angular momentum is 
\begin{equation}
\label{eq:j}
\ell = \frac{ L(t=0) - \mathcal{J}_{\rm rad}} {\nu M^2} \ ,
\end{equation}
where $\mathcal{J}_{\rm rad}$ denotes the angular momentum emitted by GWs. 
$L$ is the orbital angular momentum, a quantity that is not of direct
access in our simulations.  
Thus, we approximate $L(t=0)$ by \cite{Bernuzzi:2013rza,Dietrich:2015pxa}
\begin{equation}
 L (t=0) = J_{\rm ADM}(t=0) - S^A - S^B, 
\end{equation}
where $J_{\rm ADM}(t=0)$ is the ADM-angular momentum and $S^{A,B}$ the
spins of the NSs measured in the
initial data. We further assume that spins are approximately constant
during the evolution, cf.~Sec.~\ref{sec:Spin-Evolution}
and~\cite{Tacik:2015tja}. 

In addition to the $E_b(\ell)$ curves we consider the binding energy
and angular momentum as functions of the dimensionless parameter $x = ( M
\Omega)^{2/3}$, where $\Omega$ is the orbital frequency. The latter
can be unambiguously calculated from the simulation as \cite{Bernuzzi:2015rla}
\begin{equation}
\label{eq:MOmega}
 M \Omega = \frac{\partial E_b}{\partial \ell} \ .
\end{equation}
This quantity can be also used to characterize the postmerger
dynamics as we do in Sec.~\ref{sec:dynamics:postmerger}.

\subsubsection{Energetics: late inspiral--merger}

The $E_b(\ell)$ curves probe in a direct way the conservative dynamics of
the binary~\cite{Damour:2011fu}. In \cite{Bernuzzi:2013rza} we have
proposed a simple way of analyzing energetics during the
inspiral--merger that relies on extracting the individual contributions of the
binary interactions, i.e.~spin-orbit (SO), spin-spin (SS) and tidal
(T). Our new simulations allow us to improve that analysis by
extracting more accurately the SO and SS interaction contributions.
  
Motivated by the post-Newtonian (PN) formalism and building on
\cite{Bernuzzi:2013rza,Dietrich:2015pxa} we make the additive
ansatz for the binding energy [we omit hereafter the subscript ``$b$'']
 \begin{equation}
 E \approx E_0 + E_T + E_{SO} + E_{S^2} + E_{\SASB}, \label{eq:Eb_ansatz}
\end{equation}
where $E_0$ is a orbital (point-particle) term, $E_T$ the tidal term, $E_{SO}$ the SO term,
$E_{S^2}$ a SS term due self-coupling of the spin of the 
single star, i.e.~a change of the quadrupole moment due to the intrinsic rotation,
and $E_{\SASB}$ an SS interaction term due to the
coupling of the stars' spins. Each of the above contributions corresponds
to a term in the PN Hamiltonian. At leading order (LO) we have at 1.5PN 
\be
\label{Hso}
H_{SO} \approx \frac{2 \nu L}{r^3} S_{\rm eff}
\ee
with the effective spin 
\be
S_{\rm eff} = \left( 1 + \frac{3}{4} \frac{M ^B }{M^A}\right) \bar{S}^A + 
\left( 1 + \frac{3}{4}\frac{M^A }{M^B}\right) \bar{S}^B
\ee
and at 2PN 
\be
\label{Hs2}
H_{S^2} \approx - \frac{\nu}{2 r^3} 
\left( \frac{ C_{Q_A} M^B}{M^A}  \bar{S}_A^2 + \frac{ C_{Q_B} M^A}{M^B} \bar{S}_B^2 \right), 
\ee
and
\be
\label{Hsasb}
H_{\SASB} \approx  - \frac{\nu}{r^3} \bar{S}^A \bar{S}^B,
\ee
with $C_{Q}$ describing the quadrupole deformation due to spin, 
e.g.~\cite{Levi:2014sba} and Appendix \ref{app:PN_Ej}, and 
$\bar{S}^A = q \chi^A$, $\bar{S}^B=\chi^B/q$. 

Focusing on the equal mass configurations and applying the ansatz
above to the binding energy of each configuration, we write
\begin{eqnarray}
 E^{(00)}                  & \approx & E_0 + E_T, \\
 E^{(\uparrow \downarrow )}& \approx & E_0 + E_T + E^{(\uparrow \downarrow )}_{\SASB} 
                               + E^{(\uparrow \downarrow )}_{S^2}, \\
 E^{(\uparrow 0)}          & \approx & E_0 + E_T + E^{(\uparrow 0)}_{SO} + E^{(\uparrow 0)}_{S^2},  \\
 E^{(\uparrow \uparrow)}   & \approx & E_0 + E_T + E^{(\uparrow \uparrow)}_{SO} + 
                                 E^{(\uparrow \uparrow)}_{\SASB} +  E^{(\uparrow \uparrow)}_{S^2}.
\end{eqnarray}
We have omitted the superscript for the individual simulations 
for the $E_0$ and $E_T$ contribution since we assume that they are 
the same for all setups. 
Using the simulation data we extract each contributions as
follows. First, we consider an equal mass, 
non-spinning BBH-simulation to provide $E^{(\rm BBH)} \approx E_0$~\footnote{
The BBH $E_b(\ell)$ curve is computed with the SpEC code~\cite{SpEC}
and corresponds to SXS:BBH:0066 of the public catalog, 
see also~\cite{Blackman:2015pia}.}. Then, we use the relations
\begin{align}
&E^{(\uparrow \uparrow)}_{SO} \approx 2 E^{(\uparrow 0)}_{SO} \ , \ \ 
E_{S^2}^{(\uparrow \uparrow)} \approx 2 E_{S^2}^{(\uparrow 0)} , \\
&E_{S^2}^{(\uparrow \uparrow)} \approx E_{S^2}^{(\uparrow \downarrow)}\ , \ \ 
E_{\SASB}^{(\uparrow \uparrow)} \approx - E_{\SASB}^{(\uparrow \downarrow)} \ , 
\end{align}
that come from the LO expressions of the PN Hamiltonian above
Eqs.~\eqref{Hso}-\eqref{Hs2}-\eqref{Hsasb} and from the fact that the stars
have the same mass ($M^A=M^B$) and spin magnitudes ($S^A=S^B$).
This way, based on the five different cases, the individual
contributions read
\begin{eqnarray}
  E_{T} & \approx & E^{(00)} - E^{\rm BBH} \label{eq:ET_ex} \\
  E_{SO}^{(\uparrow \uparrow)} & \approx & - 2 E^{(00)} - E^{(\uparrow \downarrow)} + 4 E^{(\uparrow 0)} - E^{(\uparrow \uparrow)}, \\
  E_{S^2}^{(\uparrow \uparrow)} & \approx & E^{(\uparrow \downarrow)} - 2  E^{(\uparrow 0)}   + E^{(\uparrow \uparrow)}, \\
  E_{\SASB}^{(\uparrow \uparrow)} & \approx & E^{(00)} - 2  E^{(\uparrow 0)} + E^{(\uparrow \uparrow)}. \label{eq:ESS_ex}
\end{eqnarray}

All contributions are shown in Fig.~\ref{fig:Ej_H137137} for the ALF2
EOS (left) and the H4 EOS (right). For comparison we include as a
shaded region the difference between resolutions R1 and R2 for the
individual components. The plot clearly shows the repulsive
(attractive) character of the SO (tidal) interaction and quantifies
each term for a fixed value of the orbital angular
momentum. 
The plot indicates that, although poorly resolved, SS
interactions might play a role close to merger. The $E_{S^2}$ terms,
in particular, are rather large for $\ell\lesssim3.6$ and contribute
to the merger dynamics with an effect opposed to the one of the SO
interaction (note the negative sign of $E_{S^2}$ in the
plots). Summing up all the spin effects, we find that spin
contributions are of the same order as tidal effects. 
This demonstrates the importance of including spins in
analytical models of BNS.

On top of our numerical results we plot 3.5PN estimates for SO,
and SS interactions as dashed lines (see Appendix \ref{app:PN_Ej} for their explicit
expressions) and the SO effective-one-body (EOB) estimate of
\cite{Nagar:2015xqa} as dot-dashed lines. The SO term extracted from the NR data shows
significant deviations from the EOB analytical
results for $\ell\lesssim3.6-3.7$,
which correspond to GW frequencies of $M\omega_{22}\sim0.073-0.083$
(compare with merger frequencies in Tab.~\ref{tab:GWs}). The EOB model
is closer to the numerical data than the PN model, but 
underestimate (in absolute value) the magnitude of the SO term during
the last 2-3 orbits.
The PN description of SO couplings shows deviations already at
$\ell\lesssim3.8$, and it is very inaccurate
for the description of the $S^2$ SS effects\footnote{Note, 
however, that the ansatz in \eqref{eq:Eb_ansatz} might be
inaccurate at high-frequency, and that our analysis might break close to
merger. Furthermore, higher resolved simulations are needed to reduce 
the uncertainties extracting SS contributions.}.
These findings suggest that, already at the level of the Hamiltonian,
more analytical work is needed to describe the very last orbits of BNS.

Interestingly however, we note that the ``cumulative'' spin
contribution SO+SS can be reasonably approximated by the considered
EOB SO model solely
(for the considered dynamical range). The reason for this might be
that the attractive character of the SS interaction partially
``compensates'' the effect of the missing analytical 
information\footnote{Let us also point out that extracting $E_{Spin}$
is effected by smaller numerical uncertainties since only two different 
BNS configurations have to be considered, 
instead of four setups for $E_{SO}$.}.  
Let us consider the sum of all spin contributions, $E_{Spin}$, and 
assume it can be formally parametrized as the LO SO interaction (lowest order in spin)
\begin{equation}
E_{Spin} \approx 2 \nu S_{\rm eff} \mathcal{E}_{Spin} \ . 
\end{equation}
Consequently, Eq.~\eqref{eq:Eb_ansatz} simplifies to 
\begin{equation}
 E \approx E_0 + E_T + 2 \nu S_{\rm eff} \mathcal{E}_{Spin} \ , \label{eq:Ebell_ansatzSeff}
\end{equation}
and, by subtracting the non-spinning binding energy curves from the
curves for spinning configurations, we calculate $\mathcal{E}_{Spin}$.
Figure~\ref{fig:Ebell_H4} presents our results. For this analysis we
also include unequal mass configurations for which it was not possible to
extract the individual contributions to the binding energy as done
above for the equal mass cases. Notice that for an unequal mass system
also $(\uparrow \downarrow)$ configurations contain SO-interactions,
see cyan line. In the top panel we compare simulations
for different EOS. The quantity $\mathcal{E}_{Spin}$ is the same for
all simulations independent of the EOS. The bottom panel of
Fig.~\ref{fig:Ebell_H4} shows the effect of the mass ratio on
$\mathcal{E}_{Spin}$, where again up to the merger all estimates
agree. The EOB SO curve for
$\mathcal{E}_{Spin}$ are closer to the NR data in this cases than in
the one presented in Fig.~\ref{fig:Ej_H137137}.

\begin{figure*}[t]
  \includegraphics[width=1\textwidth]{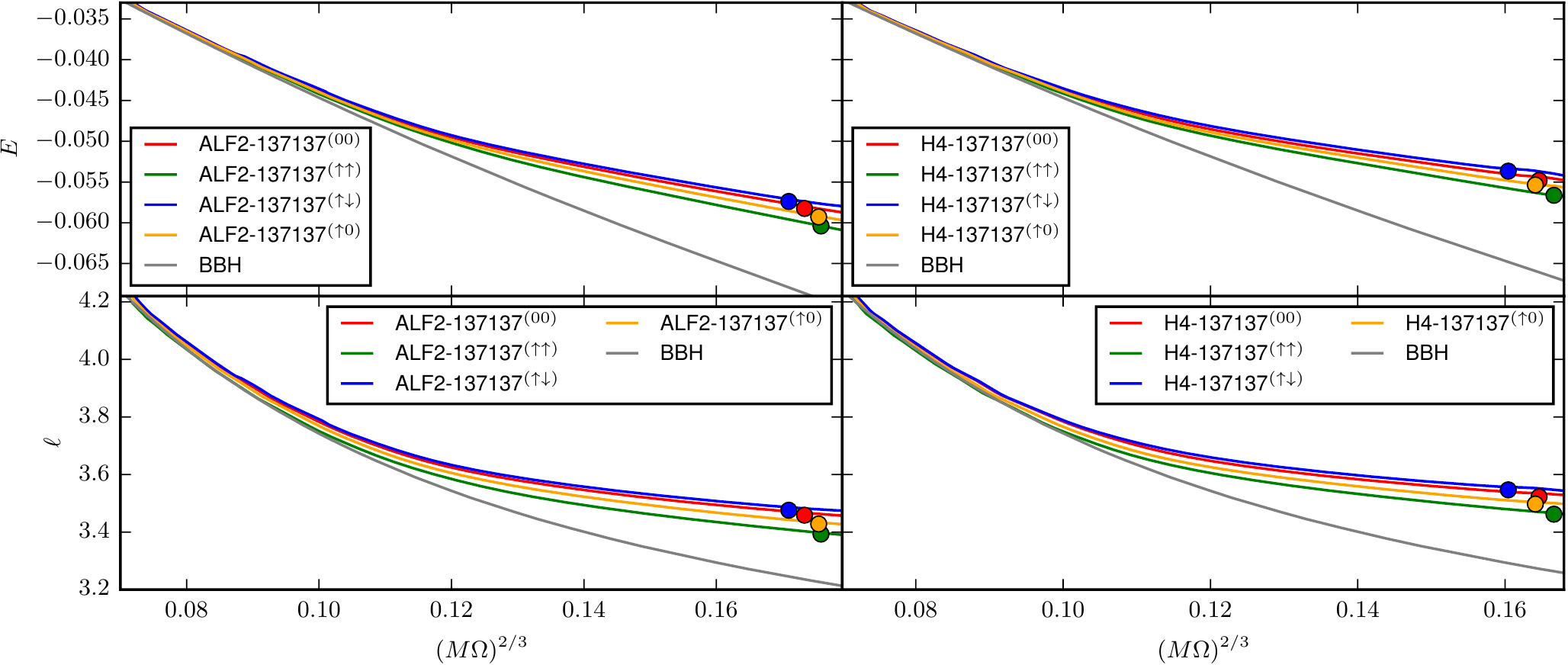}
  \caption{Binding energy (top panels) and specific angular momentum (bottom panels) as a function 
    of the PN parameter $x$. 
    The circles mark the moment of merger for all configurations. 
    We also include a non-spinning BBH configuration from the public SXS 
    catalog as in Fig.~\ref{fig:Ej_H137137}. 
    We have applied a Savitzky Golay filter on $E(x)$ and $\ell(x)$ 
    to reduce numerical noise and eccentricity oscillations. }
  \label{fig:E_ell_x}
\end{figure*} 

\begin{figure}[t]
  \includegraphics[width=0.48\textwidth]{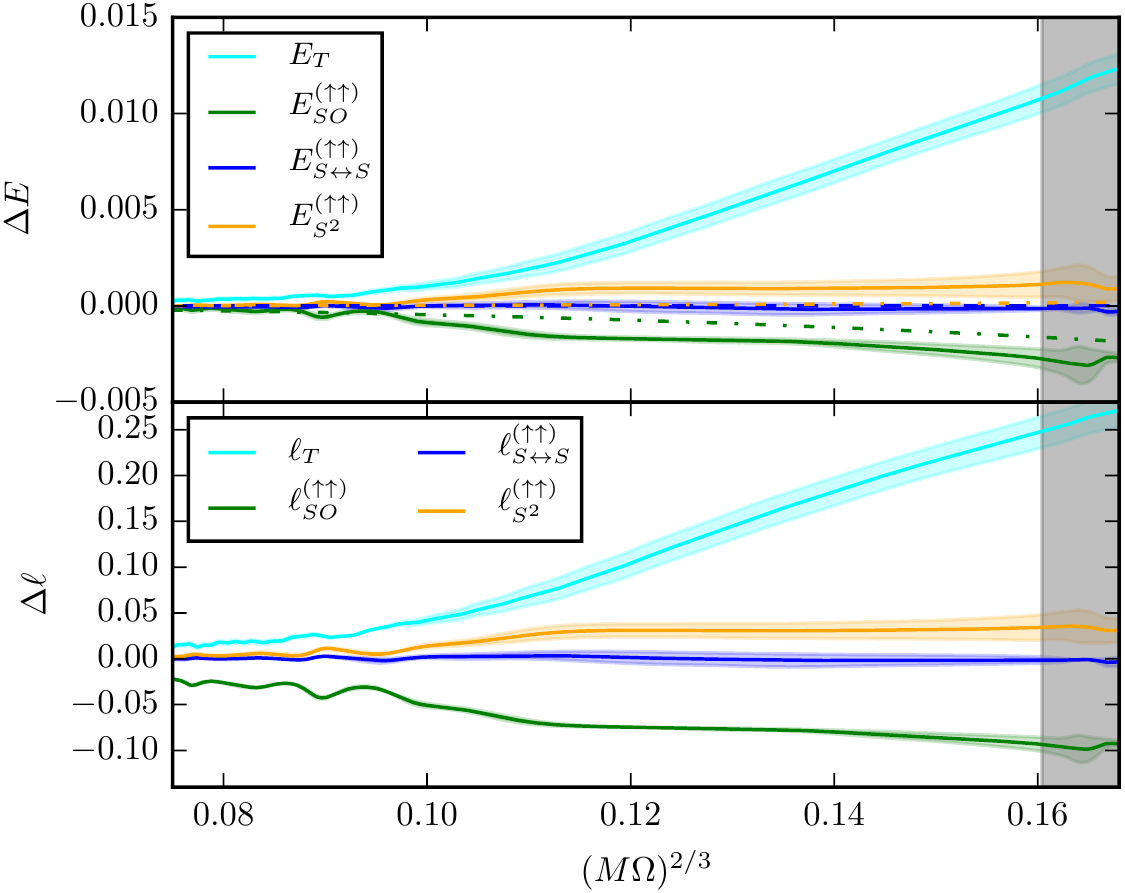}
  \caption{Individual contributions to the binding energy (top panels) and 
    specific angular momentum (bottom panels) as a function 
    of the PN parameter $x=(M \Omega)^{2/3}$ for the equal mass systems employing the H4 EOS. 
    The colored shaded regions mark the difference between resolution R2 and R1. 
    The initial oscillations are caused mostly by remaining eccentricity effects not fully 
    removed after applying the Savitzky Golay filter to $E(x)$ and $\ell(x)$. 
    We also include 3.5PN estimates for the binding energy 
    (App.~\ref{app:PN_Ej}) as dashed dotted lines. 
    The vertical gray line corresponds to the merger point of the 
    H4-137137$^{(\uparrow \downarrow)}$.}
  \label{fig:E_ell_contr}
\end{figure} 

We finally discuss the curves $E_b(x)$ and $\ell(x)$, i.e., compare
BNS energetics at fixed orbital frequency $\Omega$. Figure~\ref{fig:E_ell_x}
summarizes the equal mass results for ALF2 EOS (left) and the H4 EOS (right). 
The figure shows that once we consider systems with the 
same orbital frequency tidal contributions to 
the binding energy are larger 
than spin contributions. 
This becomes more visible in Fig.~\ref{fig:E_ell_contr} 
for which we have extracted the individual components following 
Eq.~\eqref{eq:ET_ex}-\eqref{eq:ESS_ex}. 

Figure~\ref{fig:E_ell_contr} also shows that the individual 
contributions to $E_b(x)$ and $E_b(\ell)$
have opposite signs. This can be understood by considering 
$\ell \propto \Omega r^2$ and $E \propto - r^{-1}$.

Let us first focus on tidal effects comparing a BBH and a BNS system. 
Because of the attractive nature of tidal effects
$E_{b,{\rm BBH}}>E_{b,{\rm BNS}}$ 
(but $|E_{b,{\rm BBH}}|<|E_{b,{\rm BNS}}|$) for fixed angular momentum. 
Consequently $\Omega_{\rm BBH} < \Omega_{\rm BNS}$, 
which explains the inverse ordering of 
$E_b(x)$ and $E_b(\ell)$. 
Another approach is to consider 
the $\ell(x)$ curves for a fixed frequency 
for which $\ell_{\rm BNS} > \ell_{\rm BBH}$ and 
$r_{\rm BNS} > r_{\rm BBH}$. Therefore, the system is less bound, 
i.e.~$E_{b,\rm BNS} > E_{b,\rm BBH}$, 
which is reflected in the $E(x)$ curves. 
In analogy it is possible to explain why $E_{SO}$ and other 
spin dependent contributions have opposite signs if 
$E_b(x)$ and $E_b(\ell)$ are compared.

This also shows that while $E_b(\ell)$ curves can be directly used to 
understand the effect of individual components on the conservative dynamics, the  
interpretation of $E_b(x)$ is more subtle, but will be useful for 
the phase analysis of the system presented in Sec.~\ref{sec:GW}. 

\subsubsection{Energetics: postmerger}
\label{sec:dynamics:postmerger}

\begin{figure}[t]
  \includegraphics[width=0.52\textwidth]{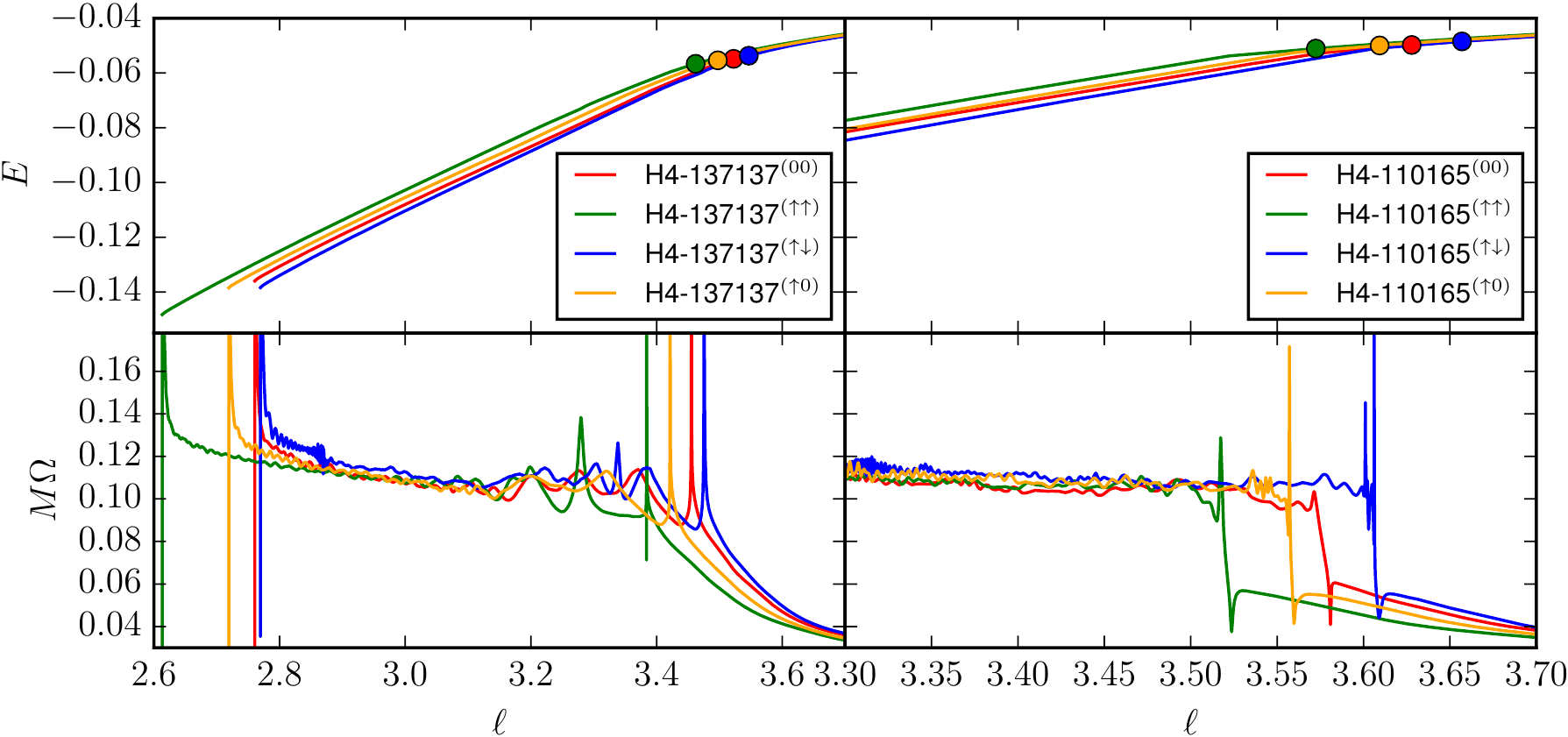}
  \caption{Binding energy vs.~specific angular momentum curves after the merger of the two neutron stars for 
    the H4 setups with mass ratio $q=1.0$ (left) and $q=1.50$ (right). 
    The merger is marked as a circle for all setups. 
    The bottom panels present the frequency $M \Omega$ estimated from the 
    binding energy curves.}
    \label{fig:Ebell_postmerger}
\end{figure}  

Binding energy vs.~specific angular momentum curves can be used also
to study post-merger dynamics \cite{Bernuzzi:2015rla}. The frequency
$M\Omega = \partial E_b/ \partial \ell$, in particular, gives the
rotation frequency of the HMNS merger remnant, and matches extremely
well half the postmerger GW frequency. Spins effects are clearly
visible at merger \cite{Bernuzzi:2013rza} but also in the postmerger
$\Omega$, especially in cases in 
which the merger remnant collapses to a black hole. 

In Fig.~\ref{fig:Ebell_postmerger} $E_b(\ell)$ and $M\Omega$
are presented for all configurations employing the H4 EOS.
When the postmerger $E_b(\ell)$ are approximately linear, the
rotational (and emission) frequency $\Omega$ remains steady for
several milliseconds.  

Comparable masses BNS remnant collapse to black hole within the
simulated times (left panels) and $\Omega$ increases continuously 
up to the collapse. The continuous evolution of
$\Omega$ (a ``post-merger chirp''
\cite{Bernuzzi:2015rla,Dietrich:2016hky}) is caused by the increasing
compactness and rotational velocity of the remnant. Spins aligned
to orbital angular momentum increase the angular momentum
support of the remnant that, therefore, collapses later in time and at
smaller values of $\ell$. The remnant of configuration
$(\uparrow\downarrow)$ has a very similar dynamics to the one of
$(00)$. 

The remnants of $q=1.5$ BNS instead do not collapse during the
simulated time. Interestingly, $\Omega$ shows a sharp jump right after
merger and then remains approximately constant. 
The jump is only present in the $q=1.5$ mass ratio setups. It
originates from the secondary star whose core ``falls'' onto 
the primary star, after a partial tidal
disruption. Consequently, the the rotational frequency of the merger
remnant experience a rapid increase over a short time.

\section{Ejecta}
\label{sec:ejecta}

  \begin{figure}[t]
  \includegraphics[width=0.45\textwidth]{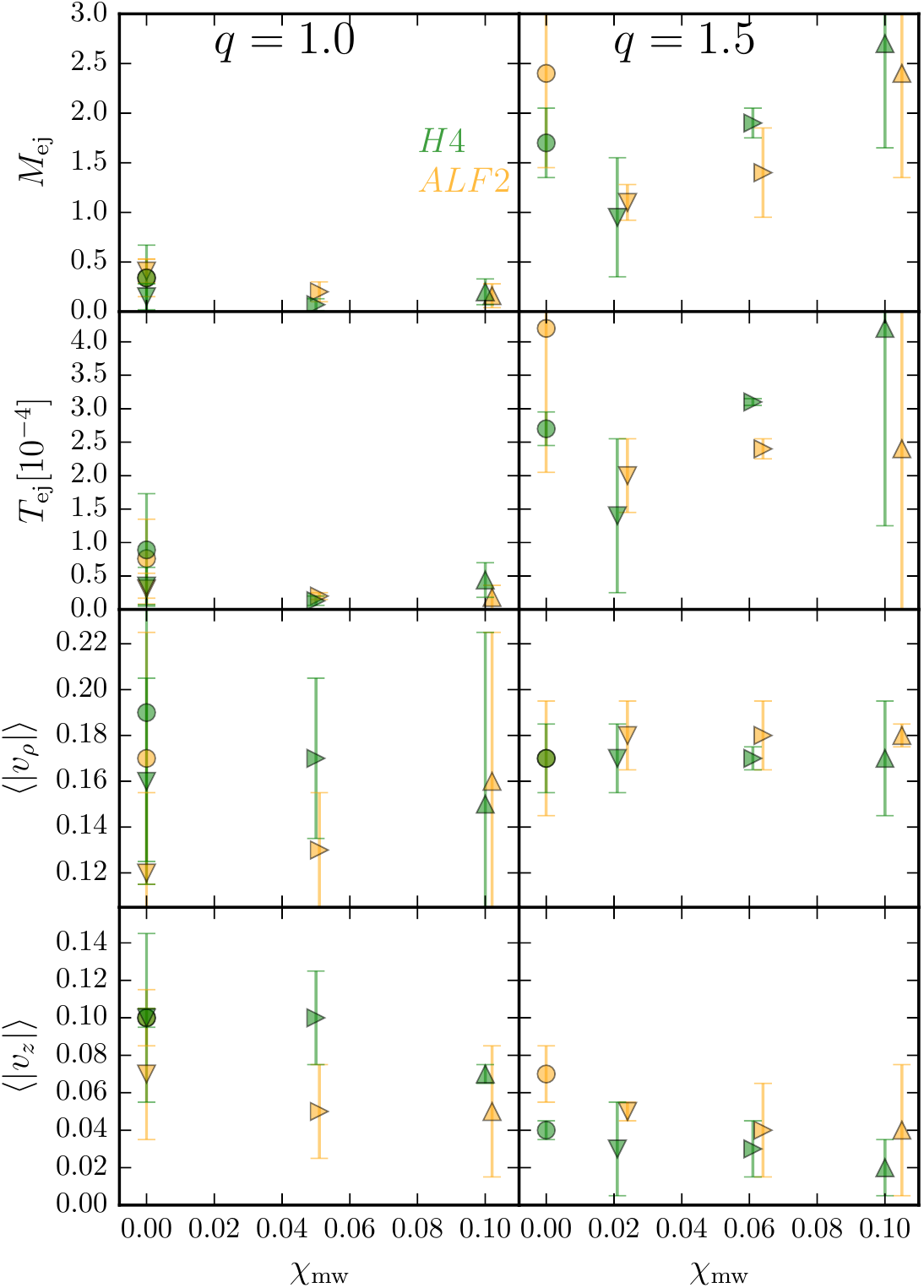}
  \caption{Mass of the ejecta (first row), kinetic energy of the ejecta (second row), 
          ejecta velocities inside the plane (third row) and orthogonal to it (fourth row)
           as a function of the spin of the configurations. 
           Green points represent data for the H4-EOS, while orange data points correspond 
           to ALF2. Left panels refer to mass ratio $q=1.00$ and right panels to mass ratio 
           $q=1.50$. 
           The markers characterize the spin of the configurations as in Fig.~\ref{fig:param}, 
           where circles correspond to $(00)$ setups, 
           down-pointing triangles to ($\uparrow \downarrow$), 
           right-pointing triangles to ($\uparrow 0$), 
           and upwards pointing triangles to ($\uparrow \uparrow$).
           The spin of the secondary star 
           influences the amount of ejected material 
           and the kinetic energy, where aligned spin leads to larger ejecta. 
           The velocity inside the orbital plane does not depend notably on $q$ or $\chi_{\rm eff}$, 
           and the velocity perpendicular to the orbital plane decreases for increasing $q$. }
  \label{fig:Mej}
  \end{figure} 

\begin{table*}[t]
  \centering    
  \caption{Ejecta properties. The columns refer to: 
    the name of the configuration, the mass of the ejecta $M_{\rm ej}$, 
    the kinetic energy of the ejecta $T_{\rm ej}$, 
    the average velocity of the ejecta inside the orbital plane $\mean{|v|}_\rho$ 
    (not necessarily pointing along the radial direction),
    the average velocity of the ejecta perpendicular to the orbital plane $\mean{|v|}_z$, 
    and the average of $v^2$ of fluid elements inside the orbital plane 
    $\mean{\bar{v}}^\rho$ and perpendicular to it $\mean{\bar{v}}^z$,
    see~Paper~I for more details. 
    Results stated in the table refer to resolution R2 and results for 
    R1 are given in brackets.}
  \begin{tabular}{l|cccccc}        
   Name    & $M_{\rm ej} \ [10^{-2}M_\odot] $    & $T_{\rm ej}\ [10^{-4}]$ & 
               $\mean{|v|}_\rho$ & $\mean{|v|}_z$  & $\mean{\bar{v}}^\rho$ & $\mean{\bar{v}}^z$ \\
     \hline
     \hline
ALF2-137137$^{00}$                  & 0.34 (0.20) & 0.76 (0.22) &  0.17 (0.12) & 0.10 (0.11) & 0.17 (0.12) & 0.22 (0.15) \\
ALF2-137137$^{\uparrow \uparrow}$   & 0.16 (0.09) & 0.18 (0.31) &  0.16 (0.10) & 0.05 (0.08) & 0.16 (0.10) & 0.14 (0.13) \\
ALF2-137137$^{\uparrow \downarrow}$ & 0.41 (0.34) & 0.31 (0.49) &  0.12 (0.15) & 0.07 (0.10) & 0.12 (0.15) & 0.12 (0.16) \\
ALF2-137137$^{\uparrow 0}$          & 0.20 (0.25) & 0.20 (0.20) &  0.13 (0.11) & 0.05 (0.07) & 0.13 (0.11) & 0.13 (0.13) \\
H4-137137$^{00}$                    & 0.34 (0.06) & 0.89 (0.10) &  0.19 (0.13) & 0.10 (0.14) & 0.19 (0.13) & 0.23 (0.22) \\
H4-137137$^{\uparrow \uparrow}$     & 0.20 (0.12) & 0.44 (0.23) &  0.15 (0.22) & 0.07 (0.07) & 0.16 (0.24) & 0.21 (0.27) \\
H4-137137$^{\uparrow \downarrow}$   & 0.15 (0.07) & 0.35 (0.12) &  0.16 (0.12) & 0.10 (0.10) & 0.17 (0.12) & 0.23 (0.20) \\
H4-137137$^{\uparrow 0}$            & 0.07 (0.06) & 0.13 (0.11) &  0.17 (0.14) & 0.10 (0.08) & 0.17 (0.14) & 0.22 (0.20) \\
     \hline
ALF2-122153$^{00}$                  & 0.75 (0.97) & 2.2 (2.1)   &  0.17 (0.09) & 0.12 (0.10) & 0.17 (0.09) & 0.23 (0.17) \\
ALF2-122153$^{\uparrow \uparrow}$   & 0.67 (0.63) & 1.4 (1.7)   &  0.16 (0.28) & 0.08 (0.06) & 0.16 (0.32) & 0.20 (0.44) \\
ALF2-122153$^{\uparrow \downarrow}$ & 0.45 (0.49) & 0.94 (0.74) &  0.15 (0.14) & 0.11 (0.09) & 0.15 (0.14) & 0.22 (0.18) \\
ALF2-122153$^{\uparrow 0}$          & 0.55 (1.9)  & 1.2 (7.8)   &  0.16 (0.17) & 0.13 (0.13) & 0.17 (0.18) & 0.21 (0.20) \\
H4-122153$^{00}$                    & 0.66 (0.88) & 1.7 (1.7)   &  0.18 (0.15) & 0.11 (0.11) & 0.18 (0.16) & 0.22 (0.28) \\
H4-122153$^{\uparrow \uparrow}$     & 0.78 (1.2)  & 1.7 (1.6)   &  0.18 (0.15) & 0.11 (0.04) & 0.18 (0.15) & 0.22 (0.16) \\
H4-122153$^{\uparrow \downarrow}$   & 0.41 (0.53) & 0.95 (1.1)  &  0.17 (0.17) & 0.09 (0.11) & 0.17 (0.18) & 0.20 (0.22) \\
H4-122153$^{\uparrow 0}$            & 0.64 (0.40) & 1.8 (1.4)   &  0.18 (0.25) & 0.08 (0.09) & 0.19 (0.28) & 0.22 (0.20) \\
   \hline
ALF2-110165$^{00}$                  & 2.4 (1.5)   & 4.2 (2.1)   &  0.17 (0.15) & 0.07 (0.08) & 0.17 (0.15) & 0.18 (0.16) \\
ALF2-110165$^{\uparrow \uparrow}$   & 2.4 (3.4)   & 4.2 (6.5)   &  0.18 (0.18) & 0.04 (0.07) & 0.18 (0.19) & 0.18 (0.17) \\
ALF2-110165$^{\uparrow \downarrow}$ & 1.1 (0.97)  & 2.0 (1.5)   &  0.18 (0.17) & 0.05 (0.05) & 0.18 (0.17) & 0.19 (0.18) \\
ALF2-110165$^{\uparrow 0}$          & 1.4 (1.8)   & 2.3 (2.5)   &  0.18 (0.17) & 0.04 (0.06) & 0.18 (0.17) & 0.19 (0.17) \\
H4-110165$^{00}$                    & 1.6 (2.0)   & 2.9 (2.9)   &  0.17 (0.16) & 0.05 (0.04) & 0.18 (0.16) & 0.17 (0.17) \\
H4-110165$^{\uparrow \uparrow}$     & 2.7 (3.7)   & 4.2 (7.1)   &  0.17 (0.19) & 0.02 (0.03) & 0.17 (0.19) & 0.15 (0.18) \\
H4-110165$^{\uparrow \downarrow}$   & 0.95 (1.5)  & 1.4 (2.5)   &  0.17 (0.18) & 0.03 (0.05) & 0.17 (0.18) & 0.17 (0.18) \\
H4-110165$^{\uparrow 0}$            & 1.9 (2.0)   & 3.1 (3.1)   &  0.17 (0.17) & 0.03 (0.04) & 0.17 (0.17) & 0.18 (0.21) \\
   \hline
   \hline
  \end{tabular}
 \label{tab:ejecta}
\end{table*}  
  
In Paper~I we have pointed out that the amount of ejected material depends significantly on the
mass-ratio where the ejecta mass increases for higher mass ratios with a
linear behavior in $q$. In large-$q$ BNSs the mass-ejection from the tidal
tail of the companion (centrifugal effect) dominates
the one originating from the cores' collision and the subsequent
shock-wave. For the same reason, stiffer EOS favor larger mass
ejection over softer EOS.
The effect of the stars' rotation (dimensionless spins $\chi\sim0.1$)
on the  dynamical ejecta are sub-dominant with respect to the mass-ratio and,
to some extend, also to varying the EOS.
We find that for configurations with
large mass ratio ($q=1.5$) the amount of ejecta is increasing from
$(\uparrow \downarrow)$ to $(\uparrow 0)$, and to $(\uparrow
\uparrow)$ due to the progressively larger angular momentum in the
tidal tail of the companion. 
We also identify a spin effects on the unbound material, as 
discussed below. 

Figure~\ref{fig:Mej} shows the most important ejecta quantities and their
dependence on the spin and the mass ratio.
We report the total ejecta mass $M_{\rm ej}$, the kinetic energy of the ejecta $T_{\rm ej}$, 
and the average velocities inside the orbital plane $\mean{|v_\rho|}$ and perpendicular to the orbital 
plane $\mean{|v_z|}$ [see Paper~I for more details]. 
The difference between resolution R2 and R1 is used as an error estimate 
and marked as an error bar. 
In the left panels results for $q=1.00$ are shown and 
results for $q=1.50$ are shown in the right panels. 
Different EOSs are colored differently: ALF2 (orange), H4 (green);
and different markers represent the different spin configurations as 
in Fig.~\ref{fig:param}.
More details about the ejecta are given in Tab.~\ref{tab:ejecta}. 

For all configurations (independent of the spin) the ejecta mass is
larger for larger mass ratios. A similar statement is true for the
kinetic energy of the ejecta (second panel of Fig.~\ref{fig:Mej}). 
The EOS variation considered here does not show significant
differences in the ejecta. Mass ejection in $q=1$ BNS 
mostly originates from the shock wave that forms during the core collision,
while in $q=1.5$ BNS mostly originates form the tidal tail.

The influence of the NS spin is smaller than the effect of the mass ratio. 
It is most visible for larger ejecta masses, i.e.~the $q=1.25,1.5$ cases, and is 
related to the spin of the companion star (less
massive NS). In a Newtonian system, 
mass ejection sets in once the fluid velocity 
is sufficiently large and the material is not bound by gravitational forces, 
i.e., once $v^2 > M_{\rm NS}/R_{\rm NS}$. The velocity of the fluid elements can be 
approximated by $v \sim v_{\rm orb} + v_{\omega}$. 
The component $v_{\rm orb}$ depends on the orbital motion and is therefore 
only indirectly effected by the spins. 
The component $v_{\omega} \approx \omega R_{\rm NS}$ is the speed of a
fluid element in the frame moving with the center of the star.
Considering the two configurations $(\uparrow \uparrow)$ and
$(\uparrow \downarrow)$, one can approximate the fluid velocity at the points farthest away from 
the center of mass as $v \sim v_{\rm orb} + |v_{\omega}|$ for $(\uparrow \uparrow)$, and 
as $v \sim v_{\rm orb} - |v_{\omega}|$ for $(\uparrow \downarrow)$ configurations. 
The criterion  $v^2 > M_{\rm NS}/R_{\rm NS}$ would be fulfilled for the
former configuration but not fulfilled for the latter. This
observation, although based on a Newtonian description, explains why
for $q\neq1$ the unbound mass increases with increasing $\chi_B$.
The observation that more material can be 
ejected for aligned configurations was also reported in~\cite{East:2015yea}
for eccentric encounters of NSBH systems using approximate initial data. 

\section{Gravitational waves}
\label{sec:GW}

In this section we discuss spin effects on the GW. In
Sec.~\ref{sec:GW:inspiral} we present, for the first time, a GW phase
analysis up to merger that quantifies the contributions of spin and
tidal interaction in the dynamical regime covered by the
simulations. 
We find that spin effects contribute to phase differences up to
$\sim5$ radians in the considered dynamical regime (for
$\chi\sim0.1$). 
In Sec.~\ref{sec:GW:postmerger} we discuss
the postmerger signal and the main emission
channels. 
We find that aligned spin configurations 
have a longer lifetime before collapse and therefore influence the 
spectral properties  of the remnant. However, 
resolving spin effects with current simulations 
in the power spectral density (PSD) of the GW signal 
is not possible. 

\begin{figure*}[t]
  \includegraphics[width=1\textwidth]{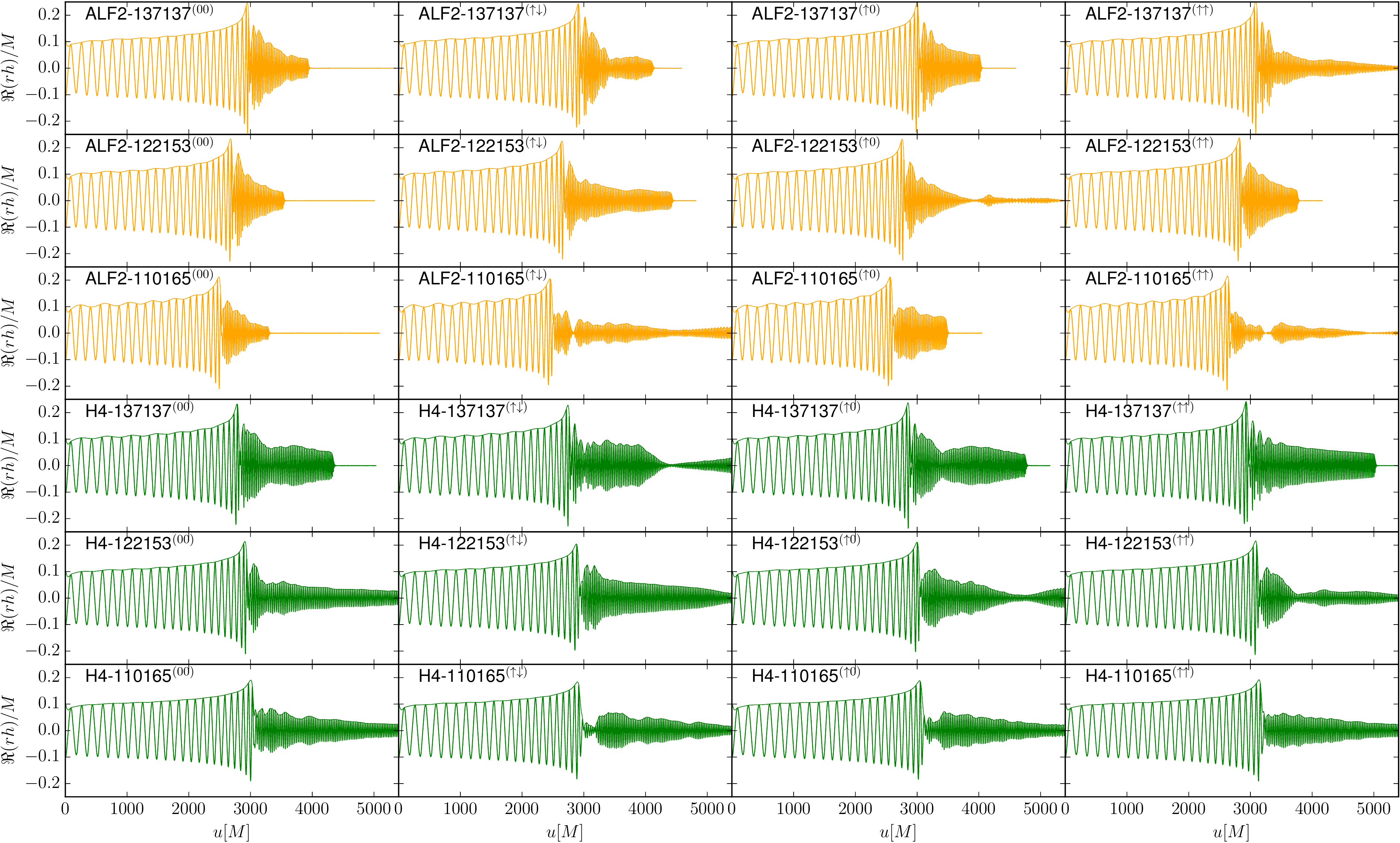}
  \caption{Gravitational wave signal for all considered configurations employing the R2 resolution. 
  Top panels (yellow lines) refer to the ALF2 EOS, bottom panels (green lines) refer the H4 EOS. 
  The mass ratio from top to bottom is: $q=1.00,q=1.25,q=1.50$. 
  The columns refer to setups: $(00)$,$(\uparrow \downarrow)$, 
$(\uparrow 0)$, $(\uparrow \uparrow)$. }
  \label{fig:GW}
\end{figure*}

Our notation follows Sec.~V of Paper I, and 
focuses on the dominant $(2,2)$ mode of the GW strain. We often use
\be
\hw := M \omega_{22}
\ee
for the dimensionless and mass-rescaled GW frequency.
GWs are plotted versus the retarded time $u$. The (real part of the)
waveforms is plotted in  Fig.~\ref{fig:GW}, as an overview of the
different signals. 
Several important quantities are listed in Tab.~\ref{tab:GWs}. 

 \begin{table*}[t]
  \centering    
  \begin{small}
  \caption{Gravitational waveform quantities. 
  The columns refer to: the name of the configuration, the number of orbits up to merger, 
  the dimensionless frequency at merger $M \omega_{mrg}$, the merger frequency in $kHz$, 
  the dominant frequencies during the post merger stage $f_1,f_2,f_3$ stated in $kHz$ and extracted 
  from the (2,1),(2,2),(3,3)-mode. 
  Results stated in the table refer to resolution R2 and results for 
  R1 are given in brackets. \label{tab:GWs}}
  \begin{tabular}{l|ccc|cccc}        
    Name & $N_\text{orb}$ &  $M\omega_{\rm mrg}$ & $f_{\rm mrg}$ & $f_1$ &  $f_{2}$ & $f_3$    \\
     \hline
     \hline
ALF2-137137$^{(00)}$                  & 11.5 (11.0) & 0.144 (0.142) & 1.69 (1.67) & 1.55 (1.46) & 2.80 (2.77) & 4.30 (4.06) \\
ALF2-137137$^{(\uparrow \downarrow)}$ & 11.3 (10.9) & 0.141 (0.138) & 1.66 (1.62) & 1.42 (1.36) & 2.77 (2.65) & 4.13 (3.78) \\                                          
ALF2-137137$^{(\uparrow 0)}$          & 11.7 (11.3) & 0.147 (0.142) & 1.73 (1.67) & 1.46 (1.35) & 2.81 (2.63) & 4.08 (3.84) \\
ALF2-137137$^{(\uparrow \uparrow)}$   & 12.0 (11.5) & 0.147 (0.144) & 1.73 (1.69) & 1.45 (1.43) & 2.75 (2.75) & 4.17 (3.99) \\                               
H4-137137$^{(00)}$                    & 10.4 (10.5) & 0.133 (0.127) & 1.56 (1.49) & 1.27 (1.38) & 2.50 (2.58) & 3.74 (3.84)  \\                                          
H4-137137$^{(\uparrow \downarrow)}$   & 10.6 (10.3) & 0.128 (0.126) & 1.50 (1.48) & 1.38 (1.37) & 2.58 (2.61) & 4.50 (3.97) \\                                                
H4-137137$^{(\uparrow 0)}$            & 11.0 (10.8) & 0.133 (0.128) & 1.56 (1.50) & 1.28 (1.35) & 2.50 (2.55) & 4.30 (4.23) \\                                   
H4-137137$^{(\uparrow \uparrow)}$     & 11.3 (11.1) & 0.136 (0.134) & 1.59 (1.57) & 1.36 (1.36) & 2.54 (2.51) &  -          \\  
\hline
ALF2-122153$^{(00)}$                  & 10.6 (10.1) & 0.133 (0.131) & 1.56 (1.57) & 1.42 (1.44) & 2.72 (2.68) & 4.11 (4.13) \\
ALF2-122153$^{(\uparrow \downarrow)}$ & 10.4 (9.9)  & 0.126 (0.123) & 1.48 (1.45) & 1.46 (1.42) & 2.73 (2.74) & 3.86 (4.17) \\                                          
ALF2-122153$^{(\uparrow 0)}$          & 10.9 (10.4) & 0.133 (0.130) & 1.56 (1.53) & 1.45 (1.38) & 2.70 (2.71) & 4.33 (4.16) \\
ALF2-122153$^{(\uparrow \uparrow)}$   & 11.2 (10.7) & 0.135 (0.133) & 1.58 (1.56) & 1.43 (1.40) & 2.75 (2.70) & 4.22 (4.10) \\                                
H4-122153$^{(00)}$                    & 10.7 (10.3) & 0.114 (0.115) & 1.34 (1.35) & 1.28 (1.24) & 2.42 (2.38) & 3.78 (3.70) \\                                      
H4-122153$^{(\uparrow \downarrow)}$   & 10.7 (10.3) & 0.108 (0.106) & 1.27 (1.25) & 1.38 (1.29) & 2.49 (2.47) & 4.26 (3.95) \\                                                
H4-122153$^{(\uparrow 0)}$            & 11.2 (10.9) & 0.112 (0.111) & 1.32 (1.30) & 1.29 (1.28) & 2.51 (2.47) & 4.07 (4.12) \\                                   
H4-122153$^{(\uparrow \uparrow)}$     & 11.6 (11.2) & 0.115 (0.114) & 1.35 (1.34) & 1.27 (1.29) & 2.49 (2.49) & 3.70 (3.79) \\
\hline
ALF2-110165$^{(00)}$                  & 9.9   (9.6) & 0.119 (0.118) & 1.40 (1.39) & 1.45 (1.32) & 2.74 (2.74) & 4.17 (4.06) \\
ALF2-110165$^{(\uparrow \downarrow)}$ & 9.7   (9.3) & 0.114 (0.113) & 1.34 (1.33) & 1.44 (1.40) & 2.82 (2.79) & 4.20 (4.24) \\                                        
ALF2-110165$^{(\uparrow 0)}$          & 10.3  (9.7) & 0.118 (0.116) & 1.39 (1.36) & 1.43 (1.33) & 2.83 (2.69) & 4.05 (4.00) \\
ALF2-110165$^{(\uparrow \uparrow)}$   & 10.6 (10.0) & 0.121 (0.120) & 1.42 (1.41) & 1.42 (1.41) & 2.80 (2.80) & 4.18 (4.24) \\                                
H4-110165$^{(00)}$                    & 11.0 (10.7) & 0.100 (0.098) & 1.17 (1.15) & 1.27 (1.24) & 2.48 (2.43) & 3.83 (3.54) \\       
H4-110165$^{(\uparrow \downarrow)}$   & 10.6 (10.1) & 0.095 (0.095) & 1.12 (1.12) & 1.29 (1.24) & 2.58 (2.50) & 3.98 (3.80) \\                                                
H4-110165$^{(\uparrow 0)}$            & 11.2 (10.8) & 0.098 (0.097) & 1.15 (1.14) & 1.29 (1.24) & 2.56 (2.53) & 3.98 (3.77) \\                                   
H4-110165$^{(\uparrow \uparrow)}$     & 11.6 (11.2) & 0.100 (0.099) & 1.17 (1.16) & 1.28 (1.27) & 2.54 (2.54) & 3.93 (3.78) \\                                              
     \hline     
     \hline  
  \end{tabular}
  \end{small}
 \end{table*}    
  
\subsection{Late-inspiral phasing}
\label{sec:GW:inspiral}

\begin{figure}[t]
  \includegraphics[width=0.5\textwidth]{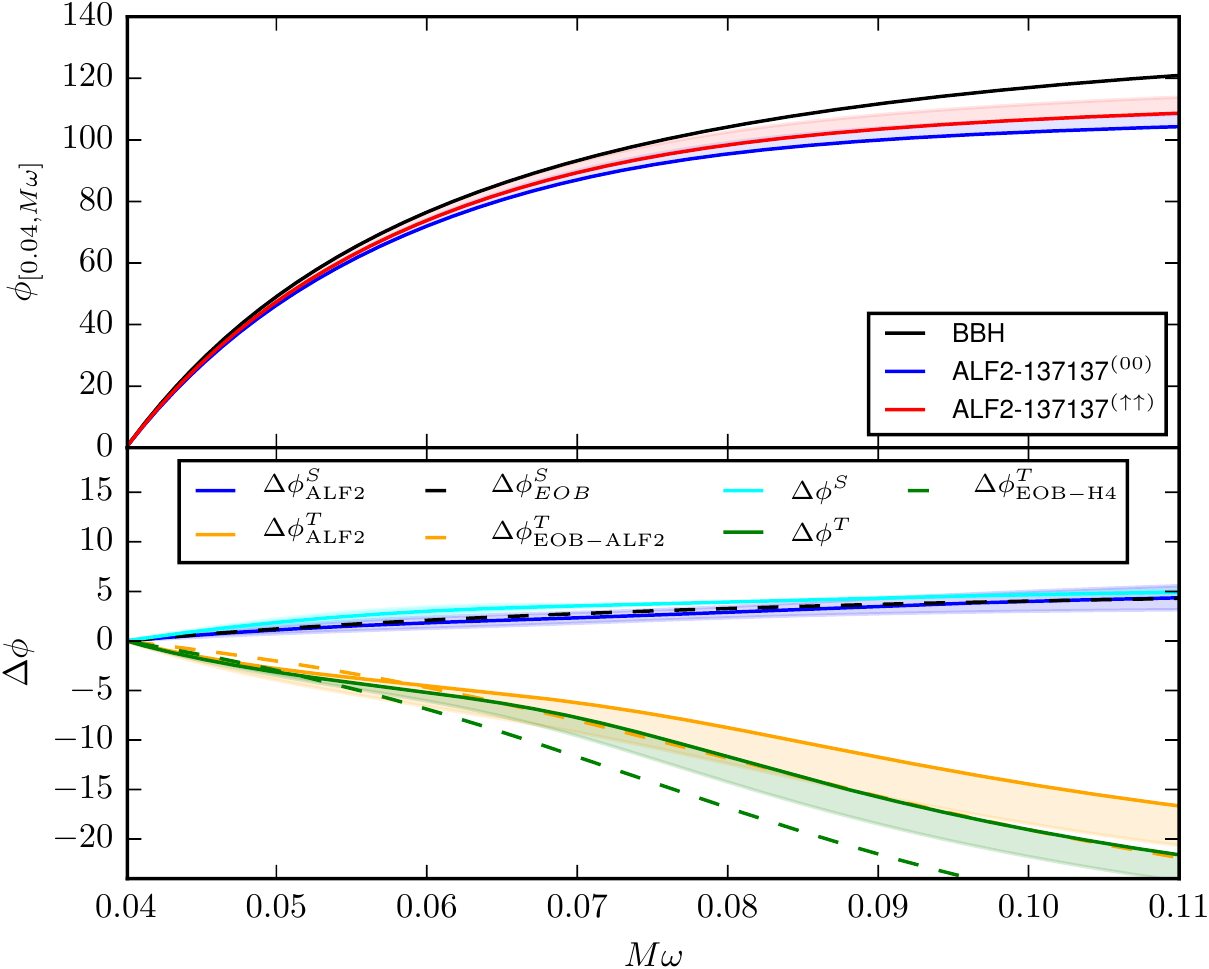}
  \caption{Top panel: $\phi(\hw)$ accumulated in $\hw\in[0.04,0.11]$ for ALF2-137137$^{(00)}$ (blue), 
           ALF2-137137$^{(\uparrow \uparrow)}$ (red), and a non-spinning, 
           equal mass BBH setup (black).
           Bottom panel: individual contributions $\Delta \phi(\hw)$. 
           We include tidal effects for ALF2 (orange) and H4 (green) EOS, as well as 
           spin effects for ALF2 (blue) and H4 (cyan). 
           We also include estimates from the EOB models of~\cite{Nagar:2015xqa}
           for the spinning contribution and~\cite{Bernuzzi:2014owa} for the tidal contribution
           as dashed lines. 
  \label{fig:Phiw}}
\end{figure}

In order to analyze the phasing of the waves we proceed as follows.
We first fit the quantity $\hw(t)$ as described in 
App.~\ref{app:omg_fit}, eliminating this way the oscillation due the
residual eccentricity in the NR data. We then integrate to obtain
$\phi(t)$ and parametrize $\phi(t(\hw))$ to obtain the phase as a function of 
the GW frequency. The integration introduces an arbitrary phase shift,
which is set to zero at an initial frequency $\hw = 0.04$. The phase
comparison is then restricted to the frequency interval $\hw \in
[0.04,0.11]$, which corresponds to physical GW frequencies
$\sim470-1292$~Hz. 

Figure~\ref{fig:Phiw} summarized our results. The upper
panel shows the phase of  ALF2-137137$^{(00)}$ (blue) and
ALF2-137137$^{(\uparrow \uparrow)}$ (red).  
The estimated uncertainty of the data is shown as a shaded region;
note that the error bar is not symmetric. The phase of a non-spinning,
equal mass BBH is included as black curve. The latter is obtained from
the EOB model of~\cite{Nagar:2015xqa}.
In the bottom panel we show the accumulated phase due to spin and
tidal interaction separately. As in the case of the
energetics, we separate the spin and 
tidal contributions to the phase by considering the difference between
the $(\uparrow \uparrow)$ and $(00)$ configuration (spin) and the
difference between the $(00)$ and the BBH configuration.  

This analysis shows that tidal effects contribute to about $15$ to
$20$ radians, accumulated in the considered frequency interval of 
$\hw \in [0.04,0.11]$. This is about 4-5 times the phase
accumulated from $10$~Hz to $\sim$ 470~Hz (i.e. from infinite
separation up to $\hw \sim0.04$) estimated with PN 
methods~\cite{Damour:2012yf}. Spin effects for $\chi \sim 0.1$ give an
accumulated phase of $\sim5$ radiants on the same frequency interval.
These results are consistent with EOB predictions included as dashed
lines.
 
Regarding the GW merger frequency (defined as the frequency at the
wave's amplitude peak), Tab.~\ref{tab:GWs} shows that BNS systems employing a stiffer EOS and/or larger
mass ratios have smaller $M\omega_{\rm mrg}$ (cf. Paper I). Spin
interactions shift the merger frequency of $\Delta M\omega\sim\pm0.005$, 
where the exact value depends on the mass ratio and EOS.

\subsection{Post-merger spectra}
\label{sec:GW:postmerger}

\begin{figure}[t]
  \includegraphics[width=0.5\textwidth]{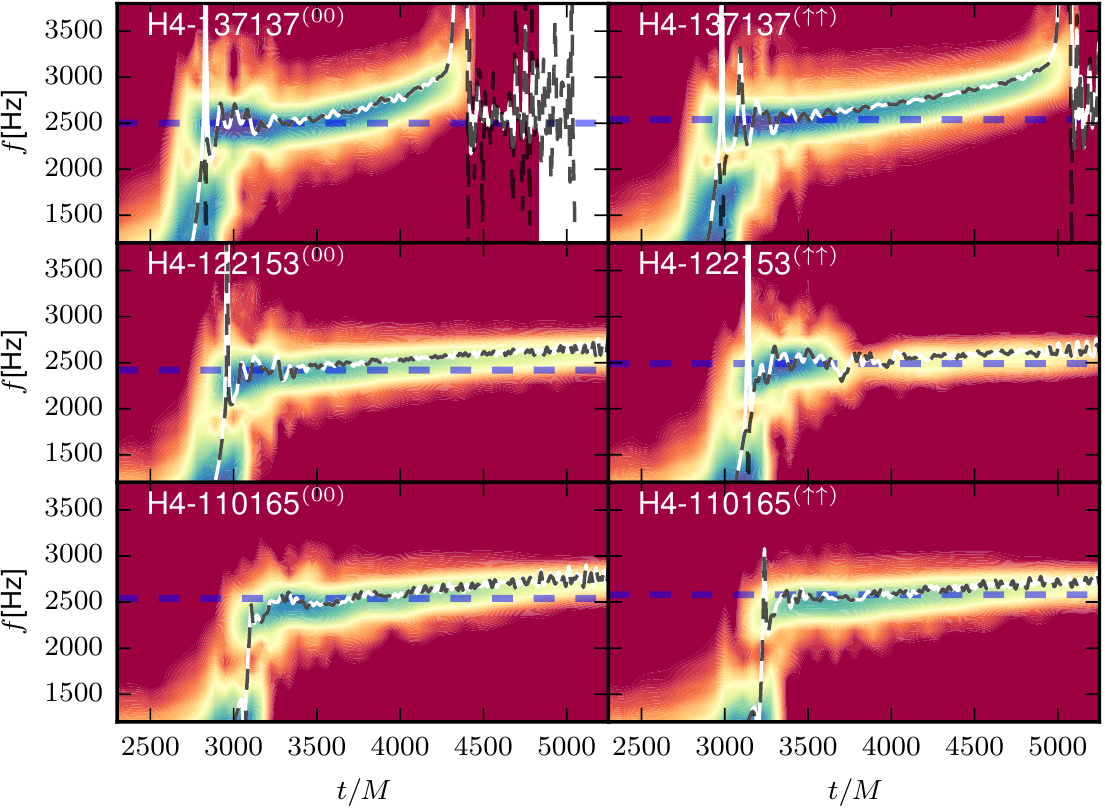}
  \caption{Spectrogram of the GW signal for simulations with H4 EOS and resolution R2. 
  The spectrogram only considers the dominant (2,2) mode. 
  Horizontal blue dashed lines refer to the $f_2$ frequency extracted from 
  the entire postmerger GW signal. Black dashed lines refer to the frequency of the (2,2)-mode
  and white lines refer to the frequency extracted from the binding 
  energy $\partial_\ell E$ (Fig.~\ref{fig:Ebell_postmerger}).
  The spectrograms show the last part of the inspiral signal (left bottom corners of the spectrograms)
  and evolution of the HMNS.}
  \label{fig:spectra}
\end{figure} 

\begin{figure}[t]
  \includegraphics[width=0.5\textwidth]{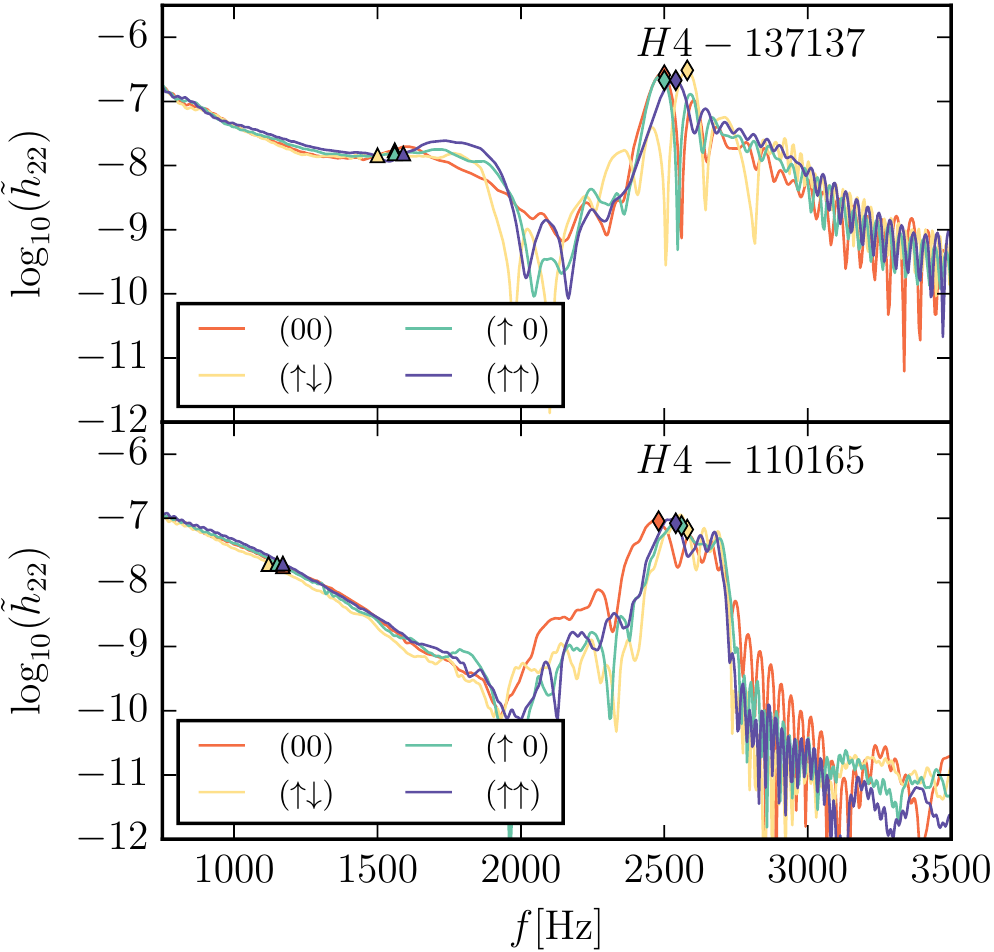}
  \caption{PSD for the H4-137137 (upper panel) and H4-110165 (lower panel) setups. 
  We mark the merger frequency with triangles and the postmerger peak frequency $f_2$
  as diamonds.}
  \label{fig:psd}
\end{figure}   

We analyze the GW spectrum of the postmerger waveform by performing a
Fourier transform of the simulation data (cf.~Sec.~V of Paper I).
Figure~\ref{fig:spectra} shows typical spectrograms of the postmerger
GW. The plot highlights the continuum character of the GW frequency,
which is especially evident in the cases in which the merger remnant is close
to collapse. This emission mirrors the dynamics discussed in
Sec. \ref{sec:dynamics:postmerger}. Due to the increasing compactness of
the remnant, the GW frequency increases until the system settles to a
stable state or collapses to a BH.  
As shown in~\cite{Bernuzzi:2015opx,Dietrich:2016hky}, however, most energy is
released shortly after the formation of the HMNS. Therefore most
of the power is at a frequency close to the one at the formation of the
merger remnant.
Spin effects are clearly distinguishable in the GW spectrum. For example, the
irrotational configuration H4-137137 
evolves faster to the collapse and has slightly lower frequency during
the postmerger then configuration $(\uparrow\uparrow)$. The frequency
drift in H4-110165 is more prominent in the irrotational $(00)$
configuration then in the $(\uparrow\uparrow)$, indicating the remnant
is closer to the threshold of radial instability (collapse).

The spectrogram plots include a horizontal blue line indicating the
``peak'' frequency $f_2$ extracted from the waveform PSD (see below). 
They also include as a white line the dynamical frequency $2 M \Omega = 2 \partial E_b/\partial \ell$ 
as computed in Sec.~\ref{sec:dynamics:postmerger} and as a black dashed line $M \omega_{22}$. 
The two frequencies remarkably agree with each other, indicating that the emission is
dominated by the non-axisymmetric $m=2$ deformation of the rotating
remnant. 

Figure~\ref{fig:psd} shows the spectrum of the signal for two exemplary
cases. Some broad peaks can be identified in the PSD, and we report
for completeness some peak frequencies in Tab.~\ref{tab:GWs}.  
As described in Paper~I, the frequencies $f_1,f_2,f_3$ 
refer to the dominant frequencies of the $(2,1)$,$(2,2)$,$(3,3)$-modes, respectively. 
Secondary peaks $f_s$ are also present,
see~\cite{Takami:2014tva,Rezzolla:2016nxn,Bauswein:2015yca,Clark:2015zxa}
for a discussion. 
As we discussed in Paper I, the $f_s$ peak at a frequency close to the
merger frequency  is basically absent for high mass ratio BNS, while a
secondary peak with slightly lower frequency as the $f_2$-frequency
becomes visible. Here we find that the secondary peak close to the merger frequency 
is enhanced for aligned spin configurations. 

In Ref.~\cite{Bernuzzi:2013rza} we reported a shift of the $f_2$
frequency of about $\sim 200$~Hz due to the spin of the NSs.  
Those simulations used higher resolutions than the ones presented here, 
but were restricted to a simple $\Gamma=2$ EOS. 
Ref.~\cite{Bauswein:2015vxa} found that for more realistic EOS but
under the assumption of conformal flatness the frequency shift is smaller.
In our new data we cannot clearly resolve frequency shifts of
$\lesssim200$~Hz, which is then to be considered an upper limit for spin
effects in BNS with the 
employed EOSs and spins $\chi\lesssim0.1$ [see also the discussion in
  Sec.~\ref{sec:dynamics:postmerger}]. 
 Nevertheless, we find in agreement with~\cite{Bernuzzi:2013rza}
 that aligned spin configurations have higher peak frequencies $f_2$. 
 For our setups the shift is only on the order of $\lesssim 50$Hz
 and thus not resolved properly. 
 Longer and higher resolved simulations will be needed for a further investigation of the 
 $f_2$-shift.
 

\section{EM counterparts}
\label{sec:EM}

Let us now discuss spin effects in possible EM counterparts in the
infrared and radio band generated from the mass ejecta.
As a consequence of the results about the dynamical ejecta, we find
that spins effects are subdominant with respect mass-ratio effect, and
more relevant the larger the unbound mass is, i.e.~for large $q$. 
However, we identify a clear trend: aligned spins 
increase the luminosity of the kilonovae and 
the radio fluency of the radio flares 
and, therefore, favor the detection of EM counterparts. 

\begin{figure}[t]
  \includegraphics[width=0.45\textwidth]{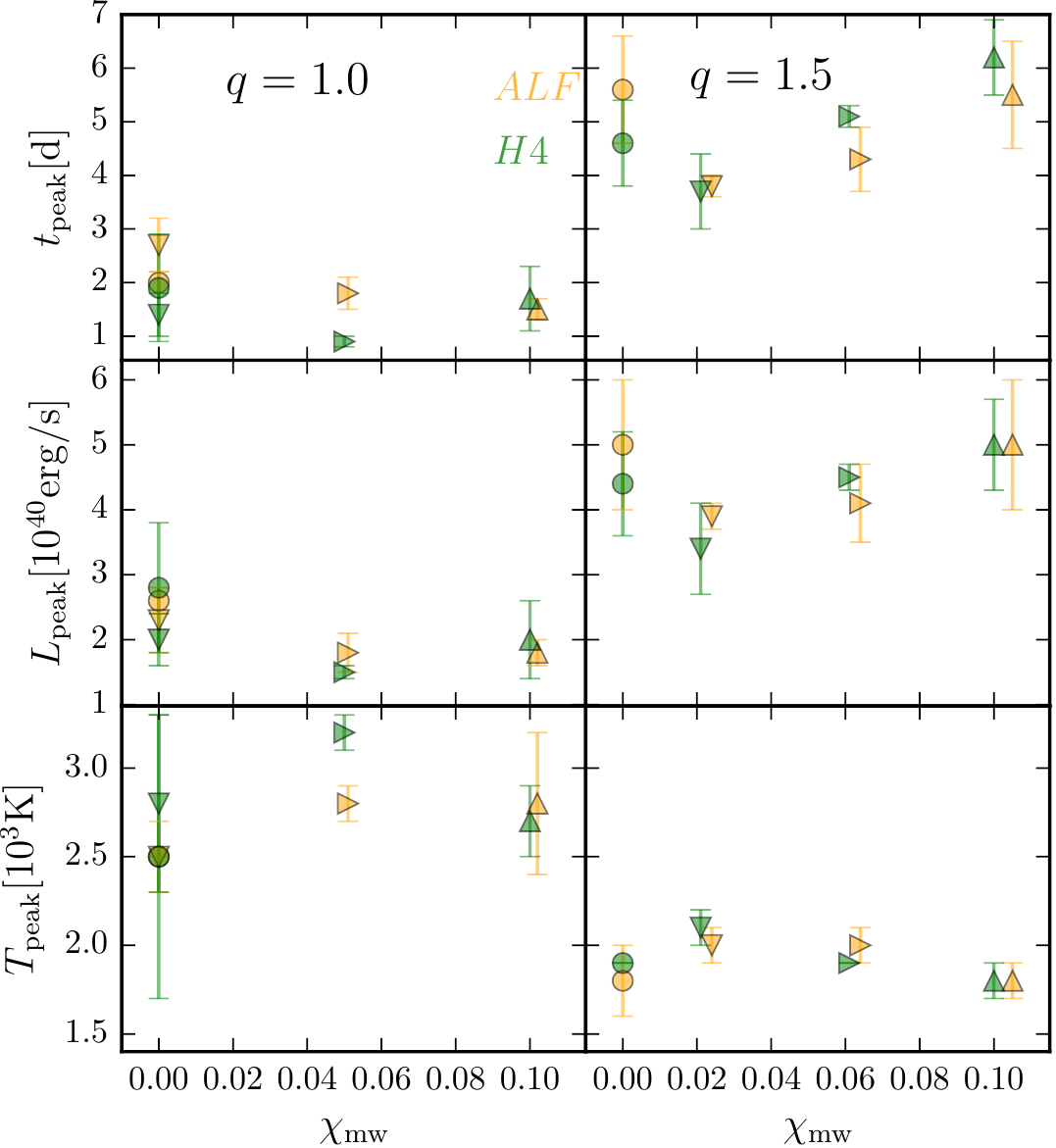}
  \caption{Peak time $t_{\rm peak}$ (top panel), peak luminosity $L_{\rm peak}$ (middle panel), 
           and peak temperature $T_{\rm peak}$ (bottom panel) of marcronovae produced 
           by the BNS mergers considered in this article as a function of the effective spin $\chi_{\rm eff}$.
           We mark different EOS with different colors: green (H4) and orange (ALF2). 
           Different markers refer to different spin configurations, see Fig.~\ref{fig:param}.}
  \label{fig:EM1}
  \end{figure}

\begin{figure}[t]
  \includegraphics[width=0.45\textwidth]{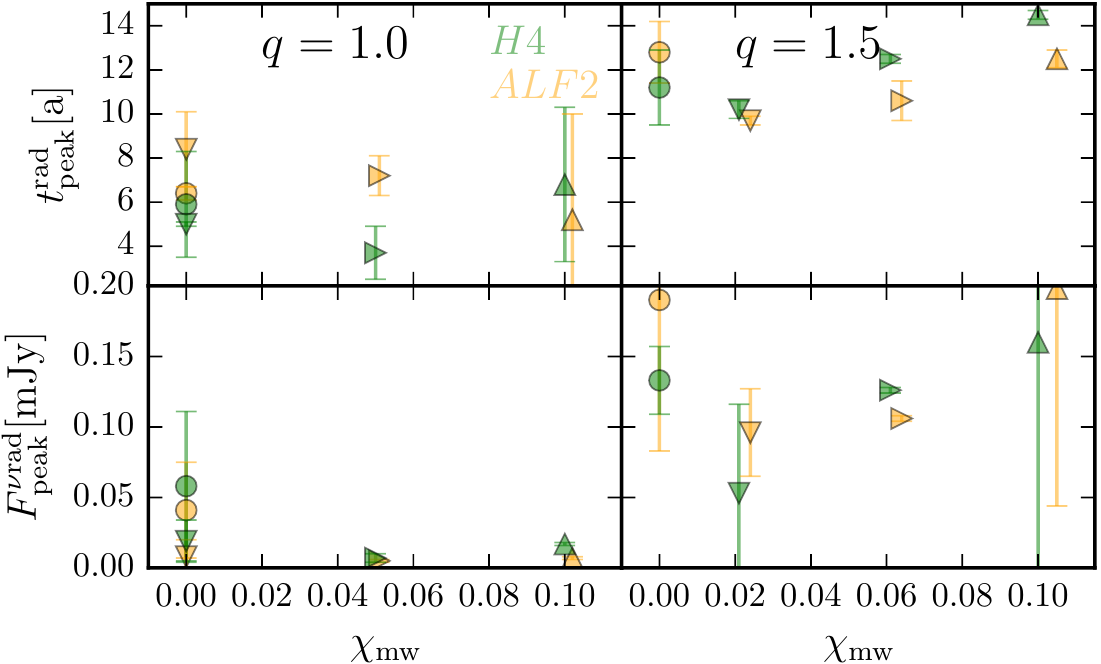}
  \caption{Peak time in the radio band $t_{\rm peak}^{\rm rad}$ (top panel) and corresponding 
           radio fluence $F^{\nu {\rm rad}}_{\rm peak}$ (bottom panel), 
           as a function of the effective spin $\chi_{\rm eff}$.
           We mark different EOS with different colors: green (H4) and orange (ALF2). 
           Different markers refer to different spin configurations, see Fig.~\ref{fig:param}.}
  \label{fig:EM2}
  \end{figure}

As in Paper~I we use the analytical model of \cite{Grossman:2013lqa}
to estimate the peak luminosity, time, and temperature  
of the marcronovae produced by the ejecta. We also use the model of~\cite{Nakar:2011cw} 
to describe radio flares peak fluxes. Our results are summarized in Tab.~\ref{tab:EM}, 
Fig.~\ref{fig:EM1}, and Fig.~\ref{fig:EM2}.

As pointed out with previous studies an increasing mass ratio delays
the luminosity peak of the kilonovae for few days, but leads to an
overall larger peak luminosity. Also, the temperature at peak
luminosity decreases for larger mass ratios. The effect of the spins
is less strong, but because of the larger ejecta mass for systems 
for which the secondary star has spin
aligned to the orbital angular momentum
we find a trend towards delayed peaks, increasing
luminosity, and decreasing temperature. 
This effect is clearly present for larger mass ratios, see
Fig.~\ref{fig:EM1}.  

\begin{figure}[t]
  \includegraphics[width=0.45\textwidth]{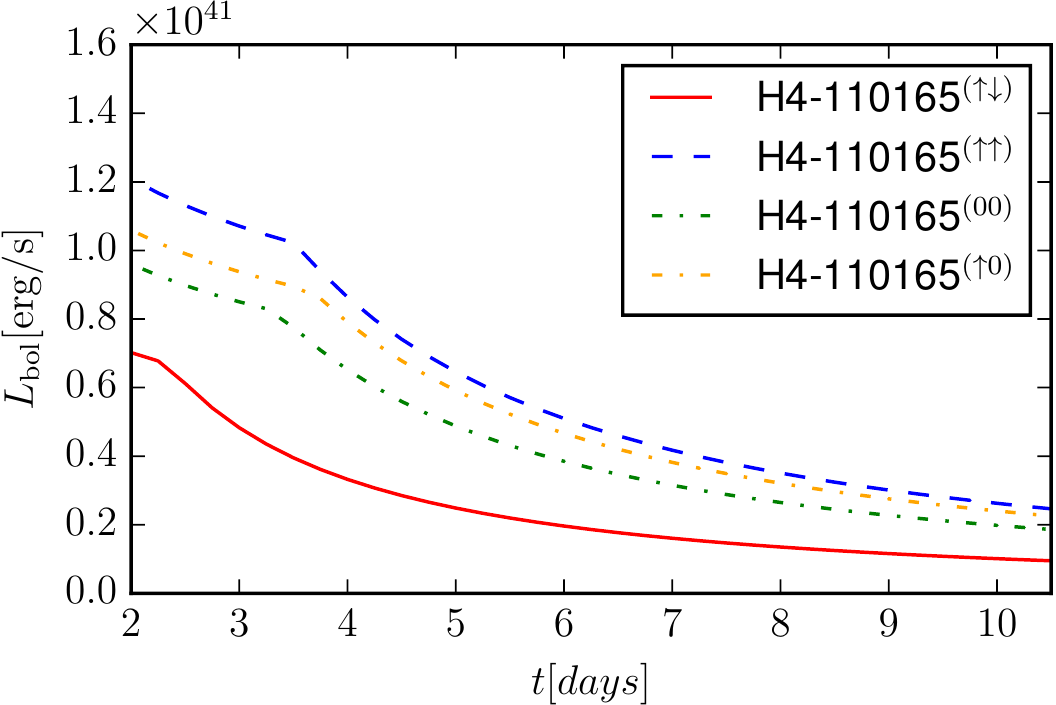}
  \caption{Bolometric luminosity for the 
           H4-110165 setups with different spin orientations. 
           The luminosities are computed following the approach of~\cite{Kawaguchi:2016ana}.}
  \label{fig:lightcurves}
  \end{figure}

We present the bolometric luminosities for the 
expected kilonova. The lightcurves are computed following the approach
of~\cite{Kawaguchi:2016ana,Dietrich:2016prep3}. 
Figure~\ref{fig:lightcurves} shows the bolometric luminosity for 
the H4-165110 setups considering different spin configurations.
Because of the larger ejecta mass for the $(\uparrow \uparrow)$ configuration
the bolometric luminosity is larger than for the other setups.
Contrary when the secondary star has antialigned spin the 
bolometric luminosity is about a factor of $\sim 2$ smaller than for
the aligned setup. 

Considering the radio flares, we find that systems with a larger mass
ratio are more likely to be detectable than equal mass setups,
Fig.~\ref{fig:EM2}. The fluency and the peak time increase  
with increasing mass ratio. Our results suggest that in cases where the less massive star has spin 
aligned to the orbital angular momentum the radio fluency increases and happens at later times.

For a more quantitative analysis higher resolution simulations and 
better models estimating the kilonova and radio burst properties are needed. 
Note also that, since our simulations are based on simulations not including microphysics 
and on simplified models, further simulations are needed to check our results. 

\begin{table}[t]
  \centering
  \setlength{\tabcolsep}{0.5pt}
  \caption{Electromagnetic Counterparts.
  The columns refer to: the name of the configuration,
                        the time in which the peak in the near infrared occurs $t_{\rm peak}$,
                        the corresponding peak luminosity $L_{\rm peak}$,
                        the temperature at this time $T_{\rm peak}$,
                        the time of peak in the radio band $t_{\rm peak}^{\rm rad}$,
                        and the corresponding radio fluence.
                        As in other tables, we present results for R2 and in brackets resolutions for R1.}
                        \begin{small} \begin{tabular}{l|ccccc}
    Name    & $t_{\rm peak}$ & $L_{\rm peak}$ & $T_{\rm peak}$ & $t^{\rm rad}_{\rm peak}$ & $F^{\nu {\rm rad}}_{\rm peak}$ \\
            & [days] & [$10^{40}\frac{\rm erg}{\rm s}$] & [$10^3$ K] & [years] & [$\mu$Jy] \\
     \hline
     \hline
ALF2-137137$^{(00)}$                   & 2.0 (1.8) & 2.6 (1.9) & 2.5 (2.7) & 6.4 (6.1)  & 41 (7) \\
ALF2-137137$^{(\uparrow \downarrow)}$  & 2.7 (2.2) & 2.3 (2.5) & 2.5 (2.5) & 8.4 (6.7)  & 8 (20) \\
ALF2-137137$^{(\uparrow 0)}$           & 1.8 (2.1) & 1.8 (1.8) & 2.8 (2.7) & 7.2 (8.1)  & 5 (4) \\
ALF2-137137$^{(\uparrow \uparrow)}$    & 1.5 (1.3) & 1.8 (1.2) & 2.8 (3.2) & 5.2 (10.0) & 7 (6) \\
H4-137137$^{(00)}$                     & 1.9 (0.9) & 2.8 (1.4) & 2.5 (3.3) & 5.9 (3.5)  & 58 (5) \\
H4-137137$^{(\uparrow \downarrow)}$    & 1.4 (1.0) & 2.0 (1.3) & 2.8 (3.3) & 5.0 (4.9)  & 19 (4) \\
H4-137137$^{(\uparrow 0)}$             & 0.9 (1.0) & 1.5 (1.3) & 3.2 (3.3) & 3.7 (4.9)  & 7 (4) \\
H4-137137$^{(\uparrow \uparrow)}$      & 1.7 (1.1) & 2.0 (2.0) & 2.7 (2.9) & 6.8 (3.3)  & 17 (18) \\
\hline
ALF2-122153$^{(00)}$                   & 2.9 (4.2) & 3.7 (2.9) & 2.2 (2.2) & 8.0 (17.6) & 139 (46) \\
ALF2-122153$^{(\uparrow \downarrow)}$  & 2.4 (2.7) & 2.8 (2.7) & 2.4 (2.4) & 7.6 (8.4)  & 44 (27) \\
ALF2-122153$^{(\uparrow 0)}$           & 2.5 (4.6) & 3.3 (5.2) & 2.3 (1.9) & 6.9 (11.9) & 74 (516) \\
ALF2-122153$^{(\uparrow \uparrow)}$    & 3.0 (2.3) & 3.2 (4.2) & 2.2 (2.2) & 8.9 (4.5)  & 64 (215) \\
H4-122153$^{(00)}$                     & 2.7 (3.4) & 3.5 (3.5) & 2.2 (2.1) & 7.3 (9.6)  & 105 (74) \\
H4-122153$^{(\uparrow \downarrow)}$    & 2.3 (2.5) & 2.8 (3.1) & 2.4 (2.3) & 7.2 (6.9)  & 48 (61) \\
H4-122153$^{(\uparrow 0)}$             & 2.8 (1.9) & 3.3 (3.5) & 2.2 (2.3) & 8.3 (4.6)  & 99 (149) \\
H4-122153$^{(\uparrow \uparrow)}$      & 3.0 (4.3) & 3.7 (3.5) & 2.1 (2.1) & 7.4 (12.0) & 108 (52) \\
\hline
ALF2-110165$^{(00)}$                   & 5.6 (4.6) & 5.0 (4.1) & 1.8 (2.0) & 12.8 (11.4) & 190 (83) \\
ALF2-110165$^{(\uparrow \downarrow)}$  & 3.8 (3.6) & 3.9 (3.6) & 2.0 (2.1) & 9.7 (9.5)   & 96 (65) \\
ALF2-110165$^{(\uparrow 0)}$           & 4.3 (4.9) & 4.1 (4.4) & 2.0 (1.9) & 10.6 (11.5) & 106 (104) \\
ALF2-110165$^{(\uparrow \uparrow)}$    & 5.5 (6.5) & 5.0 (6.0) & 1.8 (1.7) & 12.5 (12.9) & 198 (352) \\
H4-110165$^{(00)}$                     & 4.6 (5.4) & 4.4 (4.5) & 1.9 (1.9) & 11.2 (12.9) & 133 (109)\\
H4-110165$^{(\uparrow \downarrow)}$    & 3.7 (4.4) & 3.4 (4.3) & 2.1 (2.0) & 10.2 (10.6) & 53 (116) \\
H4-110165$^{(\uparrow 0)}$             & 5.1 (5.3) & 4.5 (4.5) & 1.9 (1.9) & 12.5 (12.7) & 126 (124) \\
H4-110165$^{(\uparrow \uparrow)}$      & 6.2 (6.9) & 5.0 (6.0) & 1.8 (1.7) & 14.5 (14.3) & 160 (352) \\
     \hline
     \hline
  \end{tabular}
  \end{small}
 \label{tab:EM}
\end{table}

\section{Summary}
\label{sec:conclusion}

In this article we studied the effect of the stars' rotation on equal
and unequal mass binary neutron star mergers dynamics. 
Our analysis provides a basis for future models of spin effects in
gravitational waves and electromagnetic emission.
Combined with Paper I (\cite{Dietrich:2016hky}) this work is one of the most complete
investigations of the binary neutron star parameter space available to date.

Our findings are summarized in what follows.  

\paragraph*{Energetics:}
We have considered gauge-invariant binding energy curves for both
fixed orbital angular momentum $E_b(\ell)$ and fixed orbital frequency
$E_b(x)$. The former are useful to understand the effect of individual
terms in the Hamiltionan; the latter are directly linked to the GW
phase analysis (see below).

Our new analysis of the energetics up to merger indicates that,
although the main spin effect is due to spin-orbit (SO) interactions
\cite{Bernuzzi:2013rza}, also spin-spin interaction might play a role in
the very last stage of the merger. In particular, we argue that a
self-coupling of the NS spin 
($S_A^2$ \eqref{Hs2}) 
caused by quadrupole deformation of the star due to its intrinsic rotation
contributes during the last orbits with an
attractive effect opposed to the repulsive effect of 
the SO interaction~\cite{Poisson:1997ha}.
This illustrates the importance of including also spin-spin effects in
analytical models of BNS, and poses the challenge of resolving such
effects in NR simulations. 

We note that the current best analytical representation of the SO
Hamiltonian (the effective-one-body model, EOB) shows 
some significant deviation from the NR data at small separations, see
Fig.~\ref{fig:Ej_H137137}. Curiously, comparing the EOB analytical SO model
with NR data that include also the $S^2$ interaction, we find an
effective closer agreement between the two, Fig.~\ref{fig:Ebell_H4}. 

We have used energetics and the dynamical frequency $\Omega=M^{-1}\p
E_b/\p\ell$ also to analyze the postmerger dynamics.
Spin effects are clearly visible in cases in which
the merger remnant collapses to black hole, Fig.~\ref{fig:Ebell_postmerger}.
Spins aligned to orbital angular momentum increase the angular momentum
support of the remnant, therefore, collapse happens at later times and at
smaller values of $\ell$. Spin effects on the frequency evolution of
more stable merger remnant are small and difficult to resolve. 
The $\Omega$ analysis also shows that in large-mass ratio BNS ($q\gtrsim1.5$),
the rotational frequency $\Omega$ has a sharp increase right after merger due
to the collision of the companion's core with the primary star 
(Fig.~\ref{fig:Ebell_postmerger} bottom right panel).
This fact was unnoticed in Paper I where we did
not inspect energetics. 

\paragraph*{Mass ejection:}
Spin effects in dynamical ejecta are clearly observed in unequal mass
BNS (and with large mass ratios, $q\sim1.3-1.5$), in which mass
ejection originates from the tidal tail for the companion.
Spin aligned to the orbital angular momentum favors the amount of
ejected mass because the additional angular momentum contributes to
unbinding a larger fraction of fluid elements. This effect is mostly
dependent on the spin of the companion. 

\paragraph*{Gravitational Waves:}
We presented the first analysis of spin effects in the GW phase up to
merger. Spin effects contribute to phase differences up to
$\sim5$ radians in the considered dynamical regime (and for
$\chi\sim0.1$). 
This dephasing should be compared with the $\sim 20$
radiants due to tidal effects accumulated to the BNS merger (with
respect a BBH). The hierarchy of these effects on the phase mirrors
what is observed in the energetics $E_b(x)$, Fig.~\ref{fig:E_ell_x}.
Neglecting spin effects would bias
the determination of tidal parameters in GW observations, e.g. 
\cite{Agathos:2015uaa}.
Mainly as a consequence of spin-orbit interactions,
the merger GW frequency of CRV BNS shift to higher (lower) frequencies
then irrotational BNS if a NS has aligned (antialigned) spin to the
orbital angular momentum.

The Fourier analysis of the postmerger signal indicates that spin
effects are visible in the GW spectrum (cf. Fig.~\ref{fig:psd} 
and \ref{fig:spectra}).
Some differences in the GW frequency evolution are observed in cases
the additional angular momentum support due to aligned spins
stabilizes the merger remnant for a longer time period (cf.~discussion
on spin effects on dynamics). 
Our simulations suggest that if a BNS has spin aligned to the angular momentum the
spectrum is slightly shifted to higher frequencies up to $\sim 50$~Hz
(for dimensionless spins $\chi\lesssim0.1$), but more accurate simulations
with longer postmerger evolutions will be needed to resolve the shift properly.
Furthermore, we see an effect of the spin 
on the secondary peak frequencies, where aligned configurations 
increase the enhance a secondary peak shortly after the merger frequency. 

\paragraph*{Electromagnetic counterparts:}
For the considered spin magnitudes the spin effects on the kilonovae
and radio flare properties is subdominant with respect to the mass ratio
and the EOS. Spin effects are more prominent for larger ejecta masses, where 
spin aligned (antialigned) with the orbital angular momentum increases (decreases) 
the luminosity of the kilonovae and also increases (decreases) the radio fluency of the radio flares. 
Overall we find that aligned spin BNS in combination with larger 
mass ratios favor bright electromagnetic counterparts.

\begin{acknowledgments}
  It is a pleasure to thank Bernd Br\"ugmann, Roland Haas, Tanja
  Hinderer, Nathan~K.~Johnson-McDaniel,  Harald Pfeiffer, Jan Steinhoff, Justin Vines for helpful discussions.
  We are very thankful to Alessandro Nagar for providing us with EOB 
  $E_b(\ell)$ curves and to Serguei Ossokine for providing us with the
  NR black holes binary $E_b(\ell)$.
  We thank David Radice for comments improving the manuscript. 
  W.T. was supported by the National Science Foundation under grant
  PHY-1305387.
  Computations where performed on SuperMUC at the LRZ (Munich) under 
  the project number pr48pu, Jureca (J\"ulich) 
  under the project number HPO21, Stampede 
  (Texas, XSEDE allocation - TG-PHY140019), 
  Marconi (ISCRA-B) under the project number HP10BMAB71.
\end{acknowledgments}

\appendix

\section{Quasi-local spin measures for isolated rotating NS}
\label{app:quasi_local_single}

\begin{figure}[t]
  \includegraphics[width=0.5\textwidth]{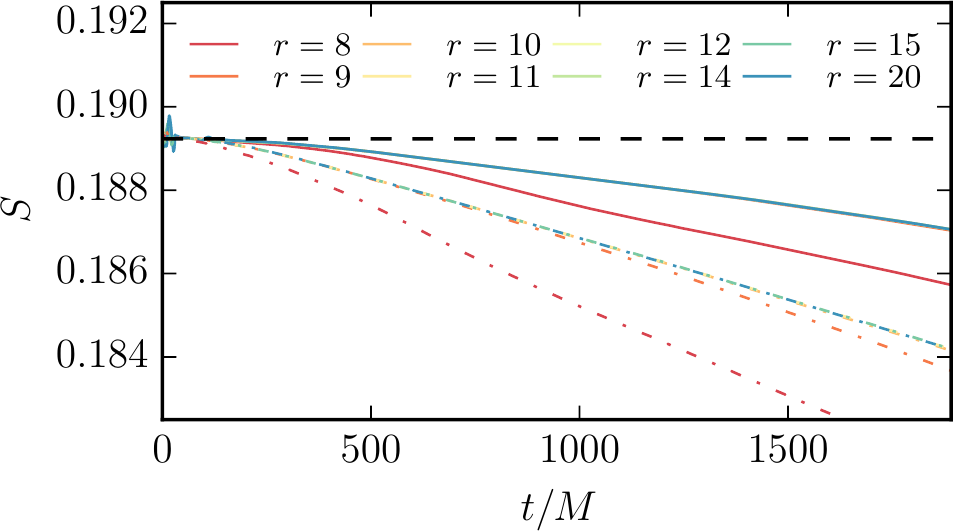}
  \caption{Quasi-local spin measure, Eq.~\eqref{eq:quasi_local_spin},
    for a single rotating NS in equilibrium. The NS has similar mass
    and spin to the components of ${\rm H4-}137137^{(\uparrow \uparrow)}$. 
    Different colors refer to different coordinate radii, where solid lines represent results for resolution R2 and 
    dashed lines for resolution R1. 
    The ADM angular momentum measured with 
    SGRID is given as a black dashed line.}
  \label{fig:quasi_local_single}
\end{figure} 

In this appendix we evaluate the accuracy of the quasi-local
measurements for isolated rotating stars in equilibrium. 
Equation~\eqref{eq:quasi_local_spin} converges to the ADM
angular momentum for $r_S\to\infty$. 
This test is important to support the interpretation of
Eq.~\eqref{eq:quasi_local_spin} as a ``spin measure''.

Figure~\ref{fig:quasi_local_single} shows our findings for a single NS
with the H4 EOS, a mass of $M=1.375$ and a dimensionless spin of
$\chi=0.10$. We compute Eq.~\eqref{eq:quasi_local_spin} for different
coordinate radii centered  around the minimum of the lapse.  
The black dashed line marks the ADM value computed by the SGRID
solver, i.e.~at $t=0$. The approximation of the ADM angular momentum by
Eq.~\eqref{eq:quasi_local_spin} improves for larger radii up to 
a point where all material is covered within the coordinate sphere.
The continuous decrease of the spin measure is caused by
numerical dissipation, and it improves at higher resolution,
cf.~difference between  solid lines (resolution R2) and dash-lines
(resolution R1).  
This test for a single NS supports our observation that the main inaccuracies of
Eq.~\eqref{eq:quasi_local_spin} for BNS is caused by the orbital motion of the two stars 
and our choice of the integration
surface.

\section{Post-Newtonian expressions for $E(\ell)$ and $E(x)$}
\label{app:PN_Ej}

We report for completeness the expressions for the 3.5PN contributions
to $E(\ell)$ and $E(x)$ used in Sec.~\ref{sec:dynamics}. 
We discard cubic terms in the spin and refer to~\cite{Levi:2014sba} for more details. 
The 4PN binding energy including higher spin terms can be found in e.g.~\cite{Levi:2015uxa,Levi:2016ofk}.  

The SO contribution is
\begin{widetext}
\begin{align}
E_{SO} (x)    &  = \nu x^{5/2} \left( \bar{S}^A + \bar{S}^B \right) \left(-\frac{4}{3} + 
                   x \left(\frac{31 \nu}{18}-4 \right)  - x^2 \left(\frac{7 \nu^2}{12} - 
                   \frac{211\nu}{8} + \frac{27}{2} \right) \right) + \nonumber \\
                                 &   + \nu x^{5/2} \left( \frac{\bar{S}^A}{q} + \bar{S}^B q \right) \left( -1 + 
                    x \left( \frac{5\nu}{3} - \frac{3}{2} \right) - x^2 \left( \frac{5 \nu^2}{8} - 
                    \frac{39\nu}{2} + \frac{27}{8}\right) \right), \\
E_{SO} (\ell) & = \frac{\nu (\bar{S}^A+\bar{S}^B)}{\ell^5} \left(2+\frac{1}{\ell^2} \left(
\frac{3 \nu}{8}+18 \right)+ \frac{1}{\ell^4}
\left(\frac{5\nu^2}{16}-27 \nu+162\right) \right) 
\nonumber \\ & 
 +   \frac{\nu (\bar{S}^A q^{-1}+\frac{\bar{S}^B} q)}{\ell^5}
 \left(\frac{3}{2}+\frac{99}{8 \ell^2}-\frac{1}{\ell^4}
 \left(\frac{195 \nu}{8}-\frac{1701}{16}\right) \right) \ . \label{Eq:SO_PN}
\end{align}
The SS $S^2$ contribution is
\begin{align}
E_{S^2}(x)    &  =  \nu ({\bar{S}^A})^2 x^3 \left( q^{-1} \left(\frac{C_{Q_A}}{2}+
             \frac{5x}{6} \left(\nu-3 \right)   + 
             \frac{ 5 x C_{Q_A}}{4} \left(\nu+1\right)\right) +  
             \nu x \left(\frac{25}{18}+\frac{5 C_{Q_A}}{3}\right)\right)
            + (A \leftrightarrow B) \ , \\
E_{S^2}(\ell) &  =  - \frac{\nu ({\bar{S}^A})^2}{\ell^6} \left (q^{-1} \left(\frac{C_{Q_A}}{2}+
             \frac{1}{8 \ell^2} \left(54 \nu+63\right)   + 
             \frac{C_{Q_A}}{4\ell^2} \left(5\nu+21\right)\right) +  
             \frac{\nu}{\ell^2}\left(\frac{65}{8}+C_{Q_A}\right)\right)
            + (A \leftrightarrow B) \ , 
\end{align}
\end{widetext}
with $C_{Q_A} \approx a \chi_A^2 M_A^3$ describing the quadrupole deformation due to spin
and $a$ depending on the EOS ranging from $\sim 1 - 10$, see e.g.~\cite{Poisson:1997ha}.
The SS $\SASB$ contribution is
\begin{align}
 E_{\SASB}(x)    & =  \nu \frac{\bar{S}^A \bar{S}^B} x^3 \left(1+ x 
  \left(\frac{5 \nu}{18}+\frac{5}{6}\right)\right) \ , \\
 E_{\SASB}(\ell) & = - \nu \frac{\bar{S}^A \bar{S}^B}{\ell^6} \left(1+\frac{1}{\ell^2}
  \left(\frac{13 \nu}{4}+\frac{69}{2}\right)\right) \ .  
\end{align}

\section{Fits of $\hat{\omega}$ and $\Qw{}$ phasing} 
\label{app:omg_fit}

\begin{figure}[t]
  \includegraphics[width=0.5\textwidth]{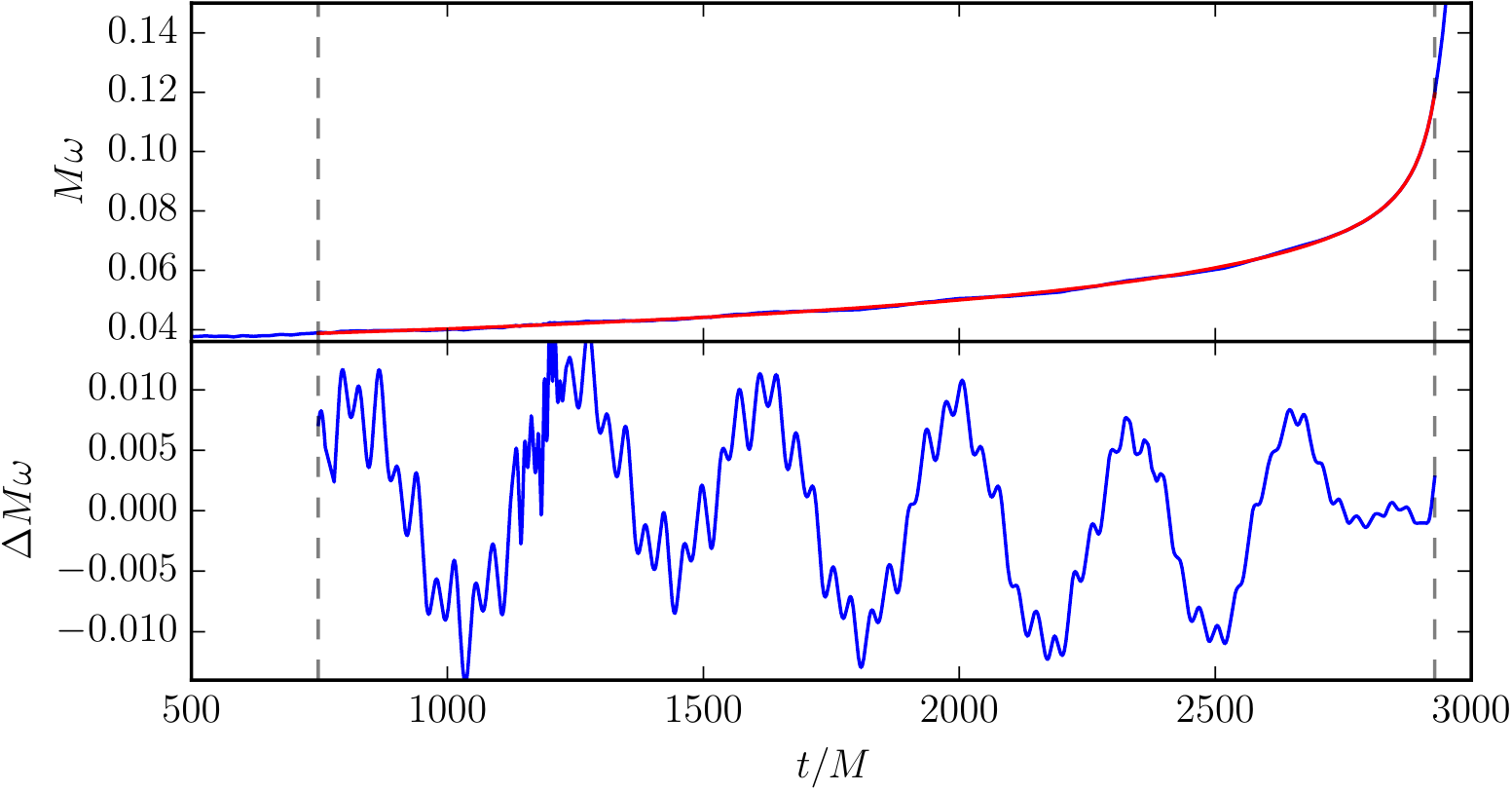}
  \caption{Example of frequency fit with template
    \eqref{eq:omg_template}. The fitting interval is
    $I_\omega=[0.039,0.129]$, covering about 18 cycles of a
    total of 23. The GW frequency at $u_\text{mrg}$ is
    $\sim0.144$ (vertical line). Data refer to ALF2 $q=1$ $M=2.75$ and no spin.} 
  \label{fig:omg_fit}
\end{figure} 

The phasing analysis of numerical data is usually complicated by 
residual eccentricity and numerical noise. To minimize these effects
we fit $\hat{\omega}(t)$ with the following template inspired by the
PN theory
\begin{equation}
\label{eq:omg_template}
\hat{\omega}(t) = \frac{1}{4}x\left( 1 + \sum_{n=1}^N c_n x^n\right)
\ ,
\end{equation}
where $x=\tau^{-3/8}$ and $\tau = ([\nu(t_c-t)/5]^2 + d^2)^{1/2}$. 
The quantities $t_c$ and $d$ are determined by the fit, using as an initial guess 
for $t_c$ the merger time and $d = (4 \hat{\omega}_{\rm
  peak})^{-8/3}$, where $\hat{\omega}_{\rm
  peak}$ is the value of the peak of $\hat{\omega}$, right after the
wave's amplitude peak. We fit on a frequency interval
$I_\omega=[\hat{\omega_1},\hat{\omega_2}]$ for $N=6$.
The fit result for an exemplary case is presented in
Fig. \ref{fig:omg_fit}. The residuals are flat and show that an 
eccentricity $\sim10^{-2}$ is ``filtered out''. 

\begin{figure}[t]
  \includegraphics[width=0.5\textwidth]{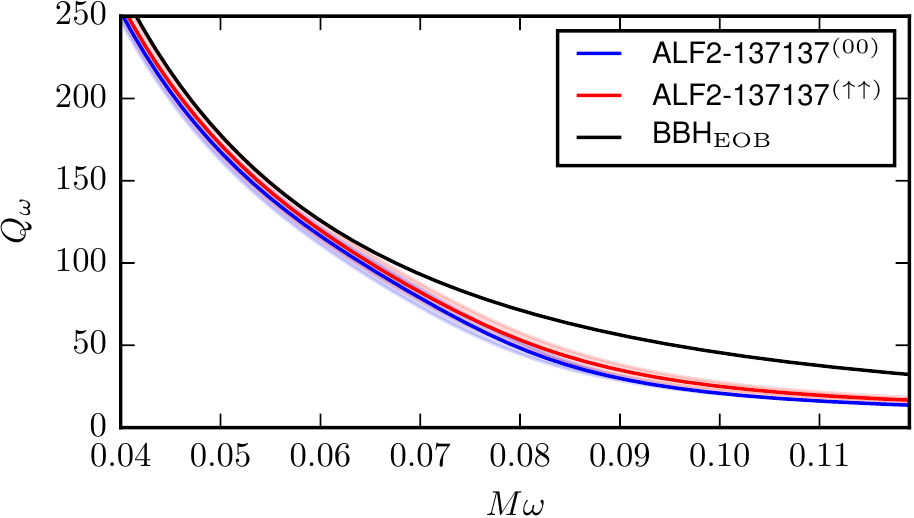}
  \caption{$\Qw(\omega)$ plot for the irrotational (blue) and $(\uparrow \uparrow)$ 
  configuration (red) employing the ALF2 EOS. $\Qw$ is computed as described in App.~\ref{app:omg_fit}. 
  We show as shaded regions the difference between different resolutions $|\Qw^{R2}-\Qw^{R1}|$. 
  As a black line we include $\Qw$ for a non-spinning, equal-mass BBH setup obtained 
  from the EOB code~\cite{Nagar:2015xqa}.
  \label{fig:Qomg_q1alf2}}
\end{figure} 

In this appendix we consider a second phasing analysis based on the quantity 
\begin{equation}
 \Qw = \frac{\hw^2}{\dot{\hw}} \ , \label{eq:Qw}
\end{equation}
where $\hw(u/M)$ and the time derivative is taken with respect the
mass-rescaled dimensionless retarded time. In PN theory $\Qw$ is an
adiabatic parameter that characterizes the validity of the stationary
phase approximation 
e.g.~\cite{Damour:2000gg}.
The phase accumulated between two different frequencies is given by
\begin{equation}
\Delta\phi(\hw_1,\hw_2) = \int^{\hw_2}_{\hw_1}
\Qw \ d \log{ (\hw)}.  \label{eq:Qw_phase}
\end{equation}
The use of $Q_\omega$ allows to perform, in principle, a phasing
analysis without manually align the waveforms in time and phase. In practice, however, the
calculation of this quantity is delicate due to numerical
inaccuracies, and a fit to the frequency needs to be used.
We verified as a further check of our fitting
procedure that the total phase computed with Eq.~\eqref{eq:Qw_phase}
is compatible with the phase of the raw data.  

In Fig.~\ref{fig:Qomg_q1alf2} we present $\Qw(\hw)$ for ALF2-137137$^{(00)}$ (blue)
and ALF2-137137$^{(\uparrow \uparrow)}$ (red) and a nonspinning $q=1$ BBH
(black, again given by the EOB model of \cite{Nagar:2015xqa}). We show as a shaded
region an uncertainty estimated as $|\Qw^{R2}-\Qw^{R1}|$. 
The largest difference between $\Qw(\hw)$ curves is due to tidal
effects, and amount to $\Delta \Qw^T\sim15-35$. From
Eq.~\eqref{eq:Qw_phase} we obtain a dephasing compatible with the
analysis of Sec.~\ref{sec:GW:inspiral}. Spin effects for $\chi\sim0.1$ and aligned
spin give a shift of about $\Delta \Qw^{Spin} \approx 5$; i.e.  
for a given frequency, aligned spins setups have a smaller $\dot{\hw}$. 
Although difficult to resolve in this plot, both tidal and spin effects are in agreement  
with predictions from the EOB model in the considered frequency interval.
A quantitative $\Qw$ analysis beyond the results stated above will
need future simulations with lower eccentricities and higher resolutions. 


\bibliography{paper20161122.bbl}

\end{document}